\pdfoutput=1
\documentclass[acmlarge, screen, nonacm]{acmart}

\AtBeginDocument{%
  }

\usepackage{graphicx}
\usepackage{subcaption}
\usepackage[table]{xcolor}
\usepackage[utf8]{inputenc}
\usepackage{newunicodechar}
\usepackage{amsmath}
\newunicodechar{₂}{$_2$}
\usepackage{float}      %
\usepackage{placeins}
\usepackage{multirow}
\usepackage{booktabs}
\usepackage{tabularx}
\usepackage{pifont}     %
\usepackage{array}      %

\newcommand{\cmark}{\ding{51}} %
\newcommand{\xmark}{\ding{55}} %
\renewcommand\footnotetextcopyrightpermission[1]{} 

\newcommand{\maybe}{%
  \cmark\hspace{-0.55em}\raisebox{0.5ex}{$x$}%
  }%

\definecolor{diagGreen}{HTML}{E3EDD5}
\definecolor{diagRed}{HTML}{F2DED6}
\definecolor{diagYellow}{HTML}{EFE8D8}
\definecolor{myGreen}{HTML}{829C49}
\definecolor{myRed}{HTML}{B46356}
\definecolor{myYellow}{HTML}{9B875D}
\definecolor{diagBlue}{HTML}{1A446C}
\definecolor{myBlack}{HTML}{000000}

\begin{document}

\title{Towards Accessible, Autonomous Photosynthesis Monitoring on Living Leaves}

\author{Qitong Li}
\email{qitongli2025@u.northwestern.edu}
\author{Xiaoyu Zheng}
\email{xiaoyuzheng2027@u.northwestern.edu}
\author{Xuanang Chen}
\email{xuanangchen2029@u.northwestern.edu}
\author{Julia Wang}
\email{juliawang2029@u.northwestern.edu}
\author{Raj Nileshbhai Dave}
\email{rajdave2027@u.northwestern.edu}
\author{Dayanara Chavez}
\email{dayanarachavez2027@u.northwestern.edu}
\author{Nirvik Prakash Iyer}
\email{nirvikiyer2029@u.northwestern.edu}
\author{Nivedita Arora}
\email{nivedita@northwestern.edu}
\affiliation{%
  \institution{Northwestern University}
  \city{Evanston}
  \state{Illinois}
  \country{United States}
}

\authorsaddresses{Authors' emails: \{qitongli2025, xiaoyuzheng2027, xuanangchen2029, juliawang2029, rajdave2027, dayanarachavez2027, nirvikiyer2029\}@u.northwestern.edu; nivedita@northwestern.edu.}

\begin{abstract}
Photosynthesis is a direct biomarker of plant health and functioning. However, measuring net photosynthesis on individual leaves requires expensive (USD 42,000 - USD 89,000) instruments and manual effort. This limits access to photosynthesis sensing and makes long-term monitoring across multiple leaves difficult. We present \textit{LeafBreath}, a $1000\times$ low-cost, autonomous sensing system that enables longitudinal monitoring of photosynthesis on living leaves over multiple days. Guided by a formative user study with 13 plant biology researchers, LeafBreath integrates an automated, leaf-attached chamber with a low-cost CO$_2$ sensor. Built using a 3D-printed clamp and thermoformed dome, the chamber periodically closes to measure CO$_2$ drawdown and reopens to refresh the air. Each chamber costs about USD 62 and operates independently, allowing multiple leaves to be monitored simultaneously without pumps or gas tubing. We characterize each chamber and validate its measurements against a research-grade gas exchange instrument (CIRAS-4) across 29 sessions ($R^2 = 0.90$). We also demonstrate monitoring of C3 and CAM photosynthesis plants, adaptability to different leaf geometries, and a light-response experiment. In a nine-day deployment, six LeafBreath chambers captured simultaneous CO$_2$ exchange from individual leaves of the same plant during drought and rewatering. Each leaf showed different patterns of nighttime CO$_2$ uptake and recovery, which a whole-plant CO$_2$ measurement instrument would obscure. To our knowledge, this is the first demonstration of such leaf-level photosynthesis dynamics using independent, low-cost autonomous chambers. Our LeafBreath work opens new opportunities for plant science, education, and data-driven agriculture.

\end{abstract}

\begin{teaserfigure}
 \includegraphics[
        width=\textwidth
    ]{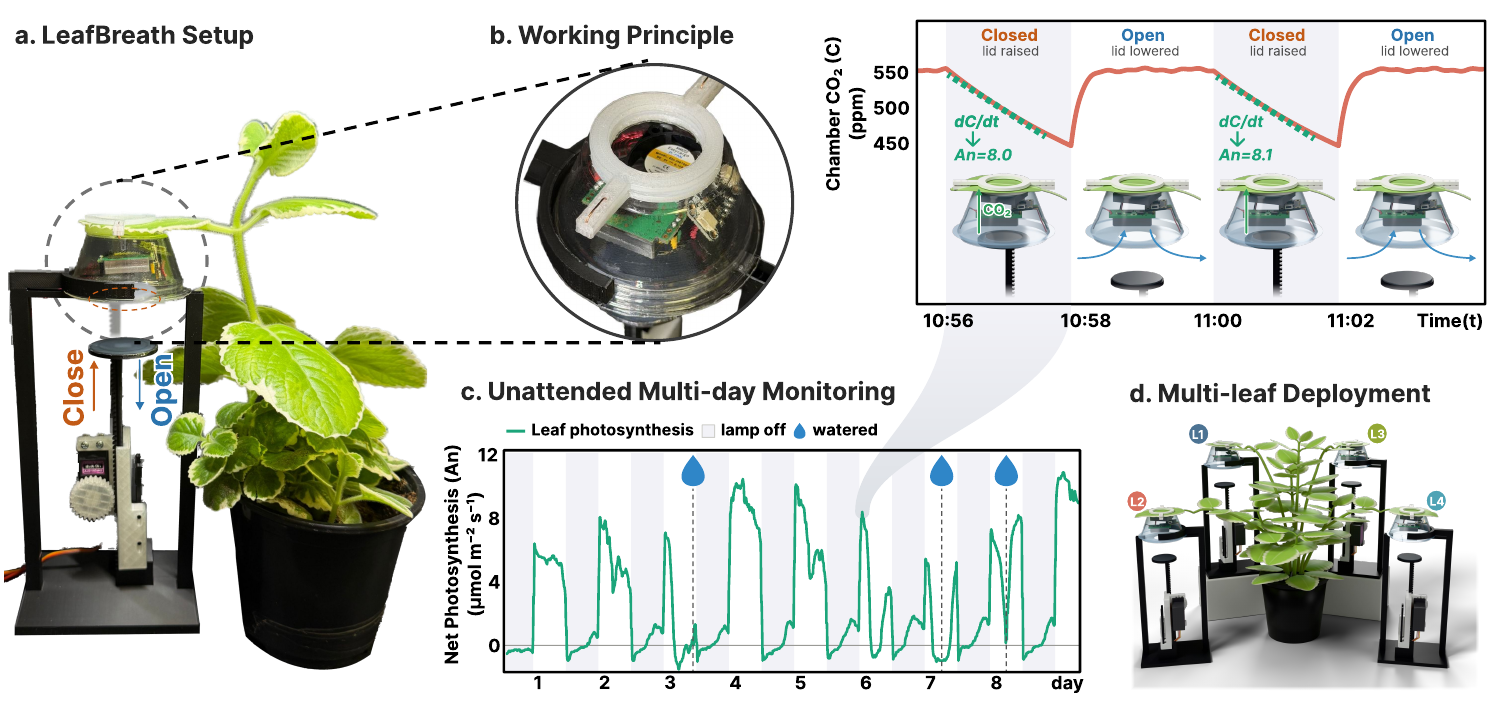}
  \caption{LeafBreath measures net photosynthesis ($A_n$) on attached leaves. (a) An automated chamber design clamped on a \textit{Coleus amboinicus} leaf. (b) A servo closes the chamber briefly. $A_n$ is computed from the slope of CO$_2$ drawdown, and the chamber reopens for refreshing. (c) $A_n$ from one leaf across the nine-day deployment (gray: lamp off, dashed: watering). (d) Multiple chambers follow multiple leaves enabling plant scientists to conduct parallel, leaf-level experiments at scale.}
  \Description{}
  \label{fig:teaser}
\end{teaserfigure}

\maketitle

\section{Introduction}
Photosynthesis is one of the Earth's most essential biological processes, producing the food we eat and the oxygen we breathe. Using light and water, the leaf converts absorbed CO$_2$ into the sugars that power all growth. Photosynthesis is also among the first processes to change when a plant is stressed, so it can reveal trouble long before leaves visibly wilt or change color~\cite{chaves2009photosynthesis,gollan1986soil,bowden1991chronology}. The photosynthetic rate is a direct biological marker of plant health. Photosynthetic rate is defined by how much CO$_2$ the leaf trades with the air~\cite{busch2024guide}. The leaf takes up CO$_2$ for photosynthesis, while it releases some through respiration. The net CO$_2$ exchange indicates whether the leaf is gaining or losing carbon. This rate, photosynthesis minus respiration, is called \textbf{net photosynthesis, or net CO$_2$ assimilation, and we write it as $A_n$.} 

Photosynthesis happens mainly in leaves, but leaves on the same plant do not contribute equally. A plant is not a single, uniform engine; it is a population of leaves of different ages and positions, each experiencing different light and water. A young leaf first imports sugar from the rest of the plant and only later exports it~\cite{turgeon1989sink}. Under stress, a plant sacrifices some leaves to protect others, moving nutrients from old leaves into young ones~\cite{lim2007leaf}. Measuring each leaf therefore reveals how the plant allocates its resources. Under drought, for example, leaves of \textit{Coleus amboinicus} switch to taking up CO$_2$ at night, and older leaves shift much more than younger ones~\cite{winter2021constitutive} (\S~\ref{sec:app-research}). During early disease, photosynthesis drops in infected leaves days before symptoms become visible, while healthy leaves continue normally~\cite{bowden1991chronology}. The standard metric plant biologists use to capture these localized shifts is net photosynthesis. A whole-plant measurement averages these differences away, and restricts both scientific discoveries and detection of early onset of plant stress. Capturing them requires sensing photosynthesis at the leaf itself, $A_n$, continuously, and across multiple leaves at the same time.

Today, measuring $A_n$ continuously means paying for an instrument that costs above 40,000 dollars (Table~\ref{tab:systems}), with an operator or a gas line at every leaf (\S~\ref{sec:methods}). Cheap sensors that attach to or sit near leaves exist, but none of them measures $A_n$ (\S~\ref{sec:signals}). A CO$_2$ sensor placed near plants can detect concentration changes associated with crop stress~\cite{tran2026rapid}. However, ambient CO$_2$ concentration alone cannot quantify photosynthetic rate or isolate the contribution of an individual leaf. Without a known volume of air over a known area of leaf, the reading cannot say how much CO$_2$ this leaf took up. 

The Ubicomp and frugal science communities have a long tradition of building low-cost alternatives to expensive instruments~\cite{cybulski2014foldscope,bhamla2017hand} and augmenting commodity hardware with modular add-ons.
For instance, a pressure sensor senses touch force when sealed in rubber or air~\cite{kim2008inflatable, tenzer2014feel}. Clips and whistles turn a phone into blood pressure and lung function monitors~\cite{BP_on_Phone_ppg, larson2012spirosmart, goel2016spirocall}. 
In each case, the sensor alone could not have made the measurement.
For a CO$_2$ sensor, enclose a known volume of air over a known area of leaf, and the change in concentration becomes $A_n$. Turning this principle into a practical system requires a design that fits different leaves, operates unattended, and produces reliable measurements. We therefore ask:

\noindent \newline \textbf{RQ1:} How do we build a low-cost, leaf-attached system that measures $A_n$ with no operator at the leaf, and that fits leaves of different sizes, shapes, and thicknesses? \newline \textbf{RQ2:} How well can a rapidly prototyped chamber measure $A_n$, and how do we characterize and validate it? \newline \textbf{RQ3:} What can continuous, multi-leaf $A_n$ records reveal that single-leaf spot measurements cannot?

To address these questions, we present \textbf{LeafBreath}, an automated leaf-attached chamber that integrates a low-cost CO$_2$ sensor (Fig.~\ref{fig:hardware}).
The clamp and chamber together enclose a known volume of air over a defined leaf area. During each measurement, a servo briefly seals the chamber while the sensor records changes in CO$_2$ concentration as the leaf takes up or releases CO$_2$. LeafBreath uses this rate of change to estimate $A_n$, then reopens the chamber to exchange air with its surroundings. This cycle repeats automatically, enabling unattended monitoring without external gas lines. Thermoforming and 3D printing allow the chamber and clamp to be adapted to different leaf geometries (Fig.~\ref{fig:fabrication}).

Our contributions are:
\begin{itemize}
\item A \textbf{formative user study} with 13 plant researchers identifying current practices, limitations, and needs for an accessible gas-exchange measurement system (\S~\ref{sec:study}). 
\item \textbf{Design of LeafBreath,} an automated leaf-attached sensing chamber that measures $A_n$ for days on multiple leaves. With a component cost of USD 62, approximately $1000\times$ lower than the price of commercial gas-exchange instruments. (\S~\ref{sec:design}, \ref{sec:application}).
\item A \textbf{characterization} of rapidly prototyped chambers and \textbf{validation} against a commercial reference gas-exchange instrument (\S~\ref{sec:char}, \ref{sec:validation}).
\item A \textbf{customizable fabrication process} for adapting chambers to leaves of different shapes, sizes, and thicknesses (\S~\ref{sec:geometry}, \ref{sec:app-adapt}).
\item \textbf{Demonstration} of physiological data collection previously limited by instrument costs: nine days of simultaneous, automated, and non-destructive photosynthesis monitoring across six leaves, capturing leaf-specific responses to drought and recovery. Additional demonstrations include multi-species measurements and a light-response experiment (\S~\ref{sec:application}).
\end{itemize}

\section{Related Work}
\label{sec:rw}
Current plant-sensing interfaces capture a wide variety of signals across multiple spatial scales. Among these approaches, LeafBreath specifically targets net photosynthesis ($A_n$) at the individual leaf scale. In this section, we review related literature across three key dimensions: the range of signals monitored in plant interfaces (\S\ref{sec:signals}), how gas-exchange is measured across different scales (\S\ref{sec:scales}), and the specific sensing methodologies used for leaf-level gas-exchange (\S\ref{sec:methods}).

\subsection{Signals from Plants}
\label{sec:signals}
To understand and interact with plants, prior human-computer interaction (HCI) and sensing systems monitor a diverse range of environmental and plant physiological signals~\cite{chang2022patterns,loh2024more}. Soil moisture, ambient light, and human care events reflect the environment the plant experiences. Meanwhile, leaf coloration~\cite{vesali2015development} and surface electrical potentials reveal plant status from the outside~\cite{steiner2017project,lee2026plants}. 
Leaf wearable sensors record temperature and humidity at the leaf surface~\cite{dong2023flexible,lee2023abaxial}, leaf water~\cite{lan2020one,peng2024wearable,li2022wearable}, emitted gases~\cite{lee2024wearable}, and chlorophyll~\cite{li2024wearable,zhang2023leaf}. 
Clip-on fluorometers record how leaves use light on many leaves for weeks~\cite{WalzMICROPAM,klueppel2026design,astashev2026development}, and the MultispeQ has carried such readings into large field campaigns~\cite{kuhlgert2016multispeq}. 
These examples showed that each piece of leaf-attached, long-term, outdoor sensing is practical.
These signals sit at different distances from photosynthesis. Photosynthesis is the leaf turning CO$_2$ into sugar, so measuring the CO$_2$ a leaf takes up measures the process directly. Soil moisture and light only describe plant surroundings. Fluorescence tracks the first step, how the leaf captures light, but captured light does not always become captured CO$_2$. Gas-exchange equipment or a model is still needed to estimate $A_n$ from these indirect signals.~\cite{han2026situ,kimura2025estimating,lee2023abaxial,li2022wearable}. \textbf{\emph{There is a need for a leaf-attached sensor that measures $A_n$ itself.}}

\subsection{Plant gas-exchange Measurement Across Scales}
\label{sec:scales}
Automated gas-exchange measurements already run across diverse scales.
Soil chambers periodically close a lid over the ground to track $\text{CO}_2$ release from the soil~\cite{forbes2023fluxbots,kobayashi2025parametric,zawilski2026sage,macagga2024validation,saccardi2025low}. Stem chambers clamp onto tree bark to measure stem respiration~\cite{brandle2019new,helm2021low}. Whole-plant and canopy chambers enclose entire plants, using motors to open and close over weeks~\cite{vaniersel2000multiple,steduto2002automated,kolling2015whole,salvatori2021low,oivukkamaki2024field,hu2023automated,alhamwi2024technical}. Many of these systems use low-cost non-dispersive infrared (NDIR) $\text{CO}_2$ sensors and run unattended for months~\cite{forbes2023fluxbots,brandle2019new,saccardi2025low,levintal2021egreenhouse}. A simple $\text{CO}_2$ sensor near crops detects plant stress~\cite{tran2026rapid}. Eddy covariance towers skip chambers entirely and track a whole field's CO$_2$ exchange from the wind~\cite{foken2012eddy}.
All these systems measure large areas or whole plants. They blend leaves, stems, and soil into a single number. None of them can track an individual leaf. 
However, a single leaf is hard to enclose. Unlike soil or tree bark, a living leaf is fragile. It cannot stay clamped, sealed, or shaded for long periods. \textbf{\emph{There is a clear need for an automated, low-cost chamber designed specifically for a single leaf.}}

\begin{table}[H]
  \caption{Leaf gas-exchange systems, by the designs in Fig.~\ref{fig:rw}. Entries describe the configurations in the cited sources, n.r.: not reported in the sources consulted.}
  \label{tab:systems}
  \footnotesize
  \setlength{\tabcolsep}{5pt}
  \renewcommand{\arraystretch}{1.25}
  \newcommand{\bad}[1]{\textcolor{myRed}{#1}}
  \newcommand{\good}[1]{\textcolor{myGreen}{#1}}
  \newcommand{\fair}[1]{\textcolor{myYellow}{#1}}
  \begin{tabularx}{\linewidth}{cllllX}
    \toprule
    Design & System & Cost & Leaves sensed & Bulk & Unattended run \\
    \midrule
    a & LI-6800~\cite{LICORBiosciences2024} & \bad{$\approx$USD 89{,}000$^1$} & \bad{1} & \bad{Bulky} & \maybe\ $\approx$8\,h per CO$_2$ cartridge$^2$ \\
    a & CIRAS-4~\cite{PPSystemsCIRAS4} & \bad{$\approx$USD 42{,}000$^1$} & \bad{1} & \bad{Bulky} & \maybe\ $\geq$12\,h per CO$_2$ cartridge$^2$ \\
    a & Leaf-mounted cuvette~\cite{frey2025ecovette} & \bad{n.r. + analyzer$^3$} & \bad{1} & \fair{Light head, bulky analyzer} & \cmark\ Multi-day \\
    b & APS~\cite{ADCAPS} & \bad{USD 71{,}300$^4$} & \good{Up to 10} & \bad{Bulky} & \cmark\ Continuous, duration n.r. \\
    c & LI-6200~\cite{licor_li6200_primer} & \bad{Discontinued} & \bad{1} & \bad{Bulky} & \xmark \\
    d & MIC-100~\cite{tanaka2022mic} & \bad{$\approx$USD 29{,}000$^5$} & \bad{1} & \bad{Bulky} & \xmark \\
    d & Low-cost chamber~\cite{takaragawa2025establishment} & \good{$\approx$USD 200$^6$} & \bad{1} & \good{Compact} & \xmark \\
    \midrule
    \rowcolor{diagGreen}
    e & \textbf{LeafBreath} & \good{$\approx$USD 62$^7$} & \good{1 per chamber} & \good{Compact} & \cmark\ 9\,d, six chambers \\
    \bottomrule
  \end{tabularx}

  \smallskip
  \raggedright\scriptsize
  $^1$Approximate 2026 prices: CIRAS-4 with PLC4 leaf cuvette (list quote), LI-6800 (manufacturer sales contact).
  $^2$One CO$_2$ cartridge lasts about 8\,h once pierced (LI-6800) and at least 12\,h (CIRAS-4), it and the scrubbing chemicals are replaced by hand. Running the CIRAS-4 from a gas cylinder instead needs an external CO$_2$ regulator (USD 1{,}200 list), but the chemicals are still replaced by hand.
  $^3$Cuvette price not reported, CO$_2$ and H$_2$O were measured with an external bench analyzer (LI-850) through tubing.
  $^4$Quote for five chambers.
  $^5$Distributor list price of the successor model MIC-100X, converted at 150\,JPY per USD.
  $^6$Materials only, excluding laptop, tools, and labor.
  $^7$Materials only, board-level parts are amortized over the chambers each board runs. Full bill of materials in Appendix~\ref{app:bom}.
\end{table}

\begin{figure}[H]
  \centering
  \includegraphics[width=0.88\linewidth]{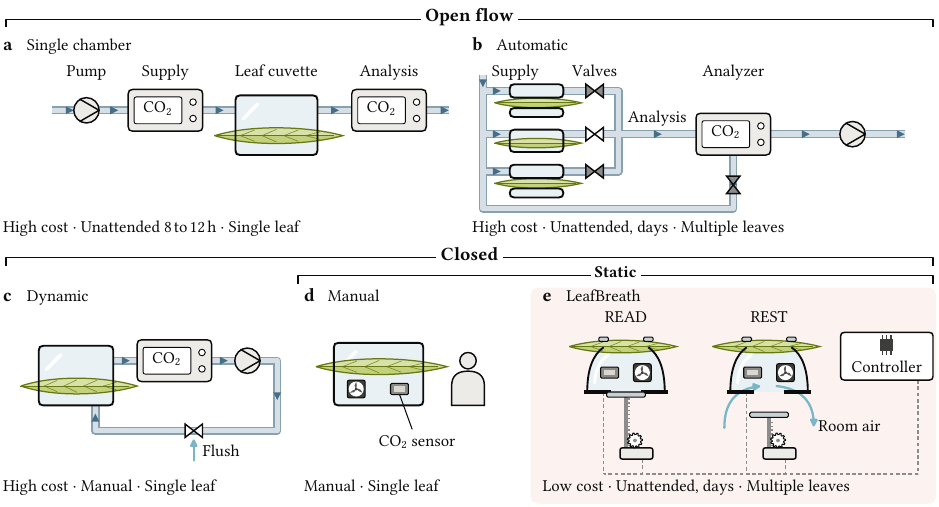}
  \caption{How leaf gas-exchange systems are built. (a) Open flow, single chamber. (b) Open flow, automatic. (c) Closed, dynamic. (d) Closed, static, manual. (e) LeafBreath: the sensor stays in the chamber and a servo closes and reopens it. Costs and demonstrated runs are in Table~\ref{tab:systems}.}
  \Description{}
  \label{fig:rw}
\end{figure}

\subsection{Leaf Gas-Exchange Instruments}
\label{sec:methods}
Fig.~\ref{fig:rw} and Table~\ref{tab:systems} compare existing tools that measure leaf gas-exchange. Each design clamps onto a leaf or encloses it in a chamber. They differ mainly in how air moves through the system.

Fig.~\ref{fig:rw}a shows the open-flow design, which is the gold standard. A pump pushes supply air through a small clamp-on chamber called a \emph{leaf cuvette}, and CO$_2$ sensors measure the CO$_2$ in the supply air and the analysis air. The difference between these two CO$_2$ sensors reveals how much CO$_2$ the leaf takes up. Consoles like the LI-6800 and CIRAS-4 also condition the air, so a leaf can be tested under chosen CO$_2$, light, and humidity~\cite{LICORBiosciences2024,PPSystemsCIRAS4,walz_gfs3000,PPSystems2024}. That control is what response experiments need, making the systems cost roughly USD 42k--89k and need an operator. However, monitoring a leaf under natural conditions does not need the control and has to be independent. Researchers mostly carry them around for quick spot checks. If left on one leaf, an unmodified unit runs for only 8 to 12 hours before its CO$_2$ cartridge and scrubber chemicals need changing by hand. Running longer takes workarounds: plumbing external air into the console~\cite{licor_li6800_gassing}, or a purpose-built leaf-wearable cuvette that stays on for days, tethered by tubing to bench CO$_2$ sensors~\cite{frey2025ecovette}.

Fig.~\ref{fig:rw}b shows the multi-leaf alternative. Its chambers open and close automatically, and valves route up to ten in turn to one central analyzer~\cite{ADCAPS,PhytoSensorPTM48A}. Every new leaf, however, requires another gas tube and lengthens the wait between readings. A five-chamber setup costs roughly USD 71{,}300 (Table~\ref{tab:systems}). \textbf{\emph{There is a clear need for an affordable tool that measures $A_n$ and scales to many leaves.}}

Closed designs need no supply air and only one analyzer. The trade-off is control. A sealed chamber cannot hold CO$_2$ or humidity steady, so closed instruments take brief readings at whatever conditions the leaf is in. They seal the chamber and watch the CO$_2$ fall. In one form, a pump loops the air past an external analyzer, and a valve flushes the loop with room air after each reading (Fig.~\ref{fig:rw}c)~\cite{licor_li6200_primer,CIDBioScience2024}. In the other, a sensor reads the CO$_2$ right at the chamber, which the operator clamps on and off by hand (Fig.~\ref{fig:rw}d)~\cite{tanaka2022mic}. Low-cost builds put the sensor inside for a few hundred dollars in parts~\cite{takaragawa2025establishment,kozlova2026development}. In every case, someone must close the chamber, watch the reading, and open it again. None has run unattended over days. \textbf{\emph{There is a need to close, read, and reopen such a chamber automatically.}}

\begin{figure}[H]
  \centering
  \includegraphics[width=0.75\linewidth]{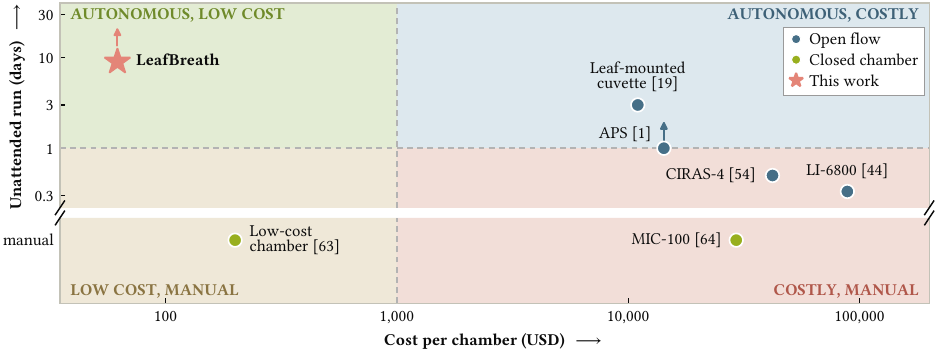}
  \caption{Leaf-scale gas-exchange systems by cost and autonomy. Each circle is a prior system for measuring leaf CO$_2$ exchange, placed by cost per chamber and by how long it can run unattended. Arrows mark a lower bound on the run: LeafBreath ran for the whole 9\,d deployment, and the automatic multi-chamber system~\cite{ADCAPS} reports no run length. The leaf-mounted cuvette~\cite{frey2025ecovette} reports no price and is placed at the estimated price of its CO$_2$ analyzer alone~\cite{cai2024evaluation}. \cite{LICORBiosciences2024,PPSystemsCIRAS4} are both expensive at USD 89,000 and USD 42,000 respectively. Existing systems are either costly or need an operator at each leaf. Of the systems shown, only LeafBreath is both unattended for days and low cost per chamber.}
  \Description{}
  \label{fig:axis}
\end{figure}

LeafBreath is that system with the operator removed (Fig.~\ref{fig:rw}e). The sensor stays inside, on the attached leaf. A servo closes the chamber briefly, the CO$_2$ slope gives $A_n$, and the chamber reopens to ambient air. Chambers run side by side, one per leaf, with no gas lines. Figure \ref{fig:axis} situates LeafBreath among existing leaf gas-exchange systems by comparing per-chamber cost and unattended runtime, highlighting its combination of low cost and multi-day autonomous monitoring. \S~\ref{sec:system} presents the design, and \S~\ref{sec:char} characterizes the leak, volume, and timing that repeated operation depends on.
\section{Formative Study: What Gas-Exchange Users Need}
\label{sec:study}
During the design of LeafBreath, we conducted an anonymous online survey to understand plant biology researchers' gas-exchange measurement needs. The study was reviewed by the university’s Institutional Review Board (IRB) and determined exempt. Our 16 questions covered current practices, instrument limitations, ideal experiments enabled by a low-cost system, and capabilities such a system must retain. The survey did not describe LeafBreath or any other new device. Of the 13 responses we received, 9 respondents had at least 6 years of experience with gas-exchange systems, their median self-reported confidence was 8 on a 9-point scale. Unless otherwise stated, all counts are out of 13.
\paragraph{\textbf{Current instruments meet core needs but constrain experimental scope.}}
Nine respondents reported that their measurement system met their need to measure different leaves on the same plant. P05 reported that their current instrument largely met their data-quality needs. However, when asked what a lower-cost system would enable, they wanted to ``include more plants and collect more data.'' In a multiple-selection question on difficulties, cost was named by twelve respondents, waiting for stable readings by seven, and operating time by four, eight reported at least one of the latter two. Physical access and sample fit also limited what could be measured: nine reported difficulty reaching leaves, including on tall plants, seven had trouble accommodating small, narrow, or irregularly shaped leaves, and four had built or modified a chamber or adapter themselves.

\paragraph{\textbf{Freed from cost, researchers would extend spatial and temporal coverage.}} %
When asked what experiments they would pursue with a low-cost system, four respondents described measurements across canopy layers or positions: P01 wanted to ``monitor different layers in the canopy,'' P09 proposed ``diurnal measurements of photosynthesis at different layers of the canopy,'' and P04 envisioned a fixed system measuring multiple canopy layers over days, weeks, or months. P07 proposed nighttime respiration or stomatal-opening measurements. These experiments share one requirement: repeated measurements of the same identified leaves over time, which with a carried instrument means revisiting each leaf by hand at every time point.

\paragraph{\textbf{The coverage researchers want is many fixed locations, frequent samples, and multi-day runs.}}
We asked how many locations on leaf or other plant organs respondents would like to monitor, how often, and for how long. Each location was a fixed area on a particular leaf or organ, and multiple selections were allowed. Nine selected more than one location, eight selected at least one sampling interval of 15 minutes or less, and five wanted to monitor the same location for more than one day, including three for more than one week. P02 presented a longer-term example: following gas-exchange before, during, and after drought for two to three months, provided that leaving the system attached did not damage the leaves or introduce measurement artifacts. %

An attached chamber also cuts the waiting seven respondents reported: the leaf stabilizes once at deployment, not before every reading, and all leaves stabilize in parallel.
Our six-leaf, nine-day deployment (\S~\ref{sec:app-research}) demonstrates the spatial and temporal coverage these requirements target.

\begin{table}[H]
  \caption{System requirements and corresponding design approaches.}
  \label{tab:requirements}
  \footnotesize
  \setlength{\tabcolsep}{5pt}
  \renewcommand{\arraystretch}{1.35}
  \begin{tabularx}{\linewidth}{p{0.30\linewidth}X}
    \toprule
    \textbf{User Requirement} & \textbf{ How LeafBreath Implements the Requirement } \\
    \midrule

    \textbf{Low cost per measured leaf}
    & Each chamber is designed to be low-cost, making it affordable to scale measurements across multiple leaves and plants (\S~\ref{sec:sensors}, Table~\ref{tab:systems}). \\

    \midrule

    \textbf{Autonomous, long-term measurement}
    & Chambers remain attached at fixed leaf locations and operate without an operator, taking measurements at intervals as short as 15\,min over multiple days (\S~\ref{sec:system} and \S~\ref{sec:readrest}). \\

    \midrule

    \textbf{Repeatable and accurate $A_n$ measurements}
    & Each hand-built chamber is characterized to assess measurement repeatability (\S~\ref{sec:char}), and measurements are compared against a reference CIRAS-4 instrument (\S~\ref{sec:aba}). \\

    \midrule

    \textbf{Minimal leaf disturbance}
    & Measurements use brief chamber closures, bounded by a drawdown limit and a timeout, to limit disturbance to the leaf. \\

    \midrule

    \textbf{Compatibility with varied leaf sizes and shapes}
    & Compact chambers require no gas lines and can operate from a USB power bank, allowing them to be mounted directly on leaves, including those high on the plant. Interchangeable clamp plates and thermoformed covers accommodate small and irregular leaves that do not fit the standard 30\,mm clamp (\S~\ref{sec:clamp} and \S~\ref{sec:app-adapt}). \\

    \bottomrule
  \end{tabularx}
\end{table}

\section{LeafBreath Chamber: Design, Fabrication and Working}
\label{sec:design}
 \subsection{System Overview}
\label{sec:system}
\begin{figure}[H]
  \centering
  \includegraphics[width=\linewidth]{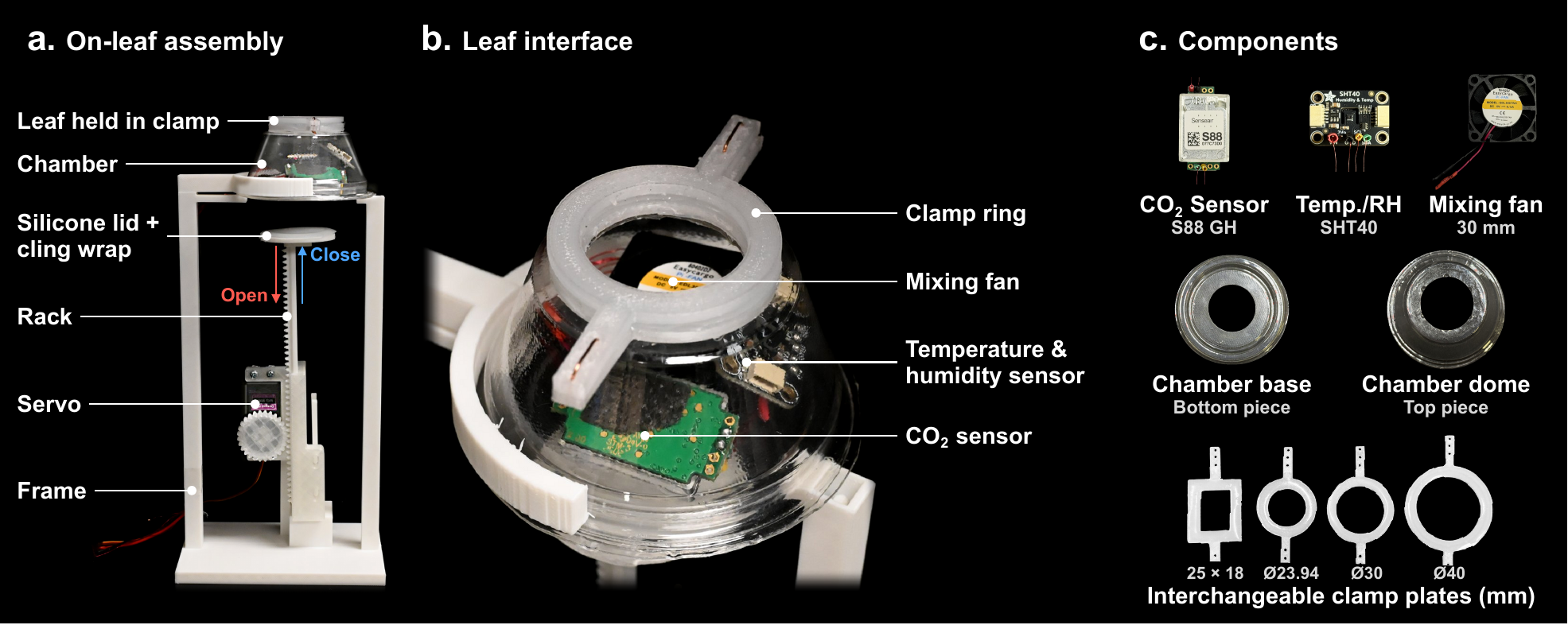}
  \caption{LeafBreath hardware. (a) The on-leaf assembly: the clamp holds the leaf over the chamber aperture, and a servo-driven rack raises a silicone lid to seal the bottom opening during each \textsc{read}. (b) Inside the chamber: the clamp ring with its cast gasket, the mixing fan, and the SHT40 and S88 GH sensors. (c) Components: sensors, fan, chamber base and dome, and interchangeable clamp plates (dimensions in mm).}
  \Description{}
  \label{fig:hardware}
\end{figure}

LeafBreath is designed to perform repeated, leaf-scale measurements of CO$_2$ gas-exchange using low-cost sensors, while minimizing disturbance to the leaf. 
The device has a transparent chamber that clamps onto the leaf. The chamber houses a CO$_2$ sensor, a temperature and humidity sensor, and a mixing fan (Fig.~\ref{fig:hardware}b).
A MG996R servo drives a rack and gear mechanism on the support frame (Fig.~\ref{fig:hardware}a) to cycle between enclosed measurements (\textsc{read}) and ambient air refresh periods (\textsc{rest}).
During each \textsc{read}, the rack raises a silicone lid covered with cling wrap to seal the bottom of the chamber dome; during \textsc{rest}, it lowers the lid to expose the chamber interior to ambient air.
The following sections explain these design choices, their trade-offs, and the measurements used to characterize them.

\subsection{Net Photosynthesis Measurement Principle}
\label{sec:theory}
LeafBreath uses the closed-chamber method. During a \textsc{read}, the chamber is sealed, and the leaf takes up CO$_2$ by photosynthesis and releases it by respiration, which changes the CO$_2$ content of the enclosed air. The sensor records the CO$_2$ mole fraction ($C$) in ppm ($\mu$mol\,mol$^{-1}$), and we fit its slope $dC/dt$.

By the ideal gas law, the chamber holds $n = PV/RT$ moles of air, where $V$ is the chamber volume excluding hardware, $P$ and $T$ are the chamber air pressure and temperature, and $R$ is the universal gas constant. The chamber therefore holds $nC$ micromoles of CO$_2$.
Taking into account only net photosynthesis, $A_n$, the amount of CO$_2$ removed per second and per unit leaf area would be
\begin{equation}
  A_n \approx -\phi\,\frac{dC}{dt}, \qquad \phi = \frac{P\,V}{R\,T\,S},
  \label{eq:simple}
\end{equation}
where $S$ is the leaf area and $\phi$ is the total amount of gas in the chamber per unit leaf area. $A_n$ is expressed in $\mu$mol\,m$^{-2}$\,s$^{-1}$, with positive values indicating photosynthesis. 

Two other processes change the chamber air during a \textsc{read}, and we correct for both.

First, the leaf adds water vapor with transpiration. The sensor's reading is the mole fraction of CO$_2$ in wet air, $C_{\mathrm{wet}}$. Transpiration lowers $C_{\mathrm{wet}}$ by adding water vapor even when no CO$_2$ is removed, and Eq.~\ref{eq:simple} would count this dilution as photosynthesis. We therefore use the dry-air mole fraction $C = C_{\mathrm{wet}}/(1-W)$ to express CO$_2$ per mole of air excluding water vapor. 
$W = e/P$ is the water-vapor mole fraction, where $e = (\mathrm{RH}/100)\,e_s(T)$ is the water vapor partial pressure, RH is relative humidity read by the temperature and humidity sensor, and $e_s$ is the saturation vapor pressure over water~\cite{buck1981new}. $n$ therefore should be representative of moles of dry air, which uses the partial pressure $P - e$.

Second, the chamber is not perfectly airtight. When the chamber CO$_2$ falls below the ambient value $C_a$, air exchange with the room draws CO$_2$ back in through leak path. This slows the drop of CO$_2$, and causes photosynthesis to be underestimated when using Eq.~\ref{eq:simple}. We measured the exchange rate $k$ of each chamber (\S~\ref{sec:char-leak}) and add the inflow $k\,(C_a - C)$.

With both corrections, the $A_n$ is
\begin{equation}
  A_n = \phi\left[-\frac{dC}{dt} + k\,(C_a - C)\right], \qquad
  \phi = \frac{(P - e)\,V}{R\,T\,S}.
  \label{eq:flux}
\end{equation}

\subsection{Sensor Selection}
\label{sec:sensors}

\paragraph{CO$_2$ Sensor.}
Placing the CO$_2$ sensor directly inside the leaf chamber imposes four key requirements: the sensor must (1)~physically fit within the compact enclosure, (2)~tolerate the humid air around a transpiring leaf, (3)~measure accurately across the sub-ambient CO$_2$ range (e.g., 200--400~ppm) as leaf photosynthesis depletes chamber CO$_2$, and (4)~sample rapidly with high resolution to capture CO$_2$ changes during brief closures. 
These constraints make package size, humidity tolerance, measurement range, and response speed central to sensor selection.

Many low-cost CO$_2$ sensors designed for indoor air quality monitoring specify an effective measurement range starting at 400~ppm, as indoor CO$_2$ rarely falls below ambient levels. 
We considered the Sensirion SCD41, which has a small footprint ($10.1 \times 10.1 \times 6.5$~mm). However, its specified accuracy applies only above 400~ppm. Moreover, it only updates once every 5~s in periodic mode with a typical response time ($T_{63}$) of 60~s, which is too slow for brief closures. 
Another candidate, the Senseair K30, measures down to 0~ppm and has been integrated into a closed leaf chamber~\cite{takaragawa2025establishment}, but its $51 \times 58 \times 12$~mm footprint requires substantially more space, limiting the flexibility of the chamber design.

We therefore selected the Senseair S88 GH NDIR sensor~\cite{senseair_s88gh}. 
It combines a compact $33.9 \times 19.8 \times 8.7$~mm package with a full 0--20,000~ppm calibrated range, operates reliably up to 95\% relative humidity, and provides a faster response ($T_{63} < 30$~s). 
We record its digital CO$_2$ output at 1~Hz with 1~ppm resolution. 
We also disable its built-in automatic baseline correction (ABC) because photosynthetic drawdowns regularly take chamber CO$_2$ below 400~ppm, which would otherwise corrupt the sensor's baseline.

\paragraph{Air Temperature and Humidity Sensor.} 
A Sensirion SHT40 module measures chamber air temperature ($T$) and relative humidity at 1~Hz for the dry-air correction in each chamber (Eq.~\ref{eq:flux}). 
Although our electronics support an optional fine-wire thermocouple for leaf temperature, a dedicated leaf temperature sensor is omitted from the standard configuration because it is not required to calculate the net exchange of CO$_2$.

\begin{figure}[H]
  \centering
  \includegraphics[width=\linewidth]{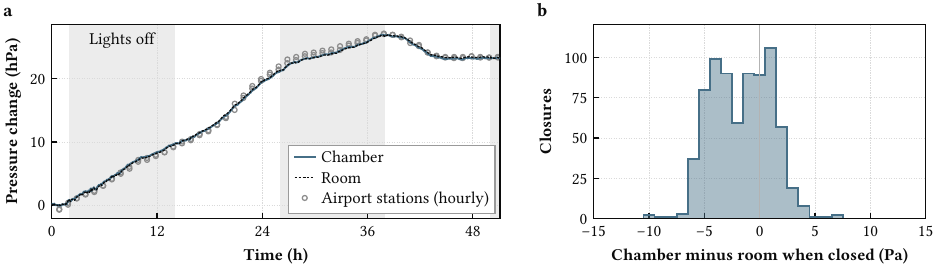}
  \caption{Air pressure in a closed chamber. (a) Pressure change over two days in the chamber (BMP280), in the room, and at two nearby airport stations, each offset to the room sensor, shaded: grow lights off. (b) Mean chamber minus room pressure during each of 745 closures, relative to the minute before closing.}
  \Description{Two panels. (a) Over about 50 hours the pressure rises by about 27 hectopascals and then falls back to about 23, the chamber, room and airport traces lie on top of each other. (b) A histogram of the pressure change per closure, between about minus 6 and plus 3 pascals for most closures and never beyond 11 pascals.}
  \label{fig:pressure}
\end{figure}

\paragraph{Air Pressure Sensor.}
\label{sec:pressure}
$A_n$ scales with $P$ (Eq.~\ref{eq:flux}). Room pressure varied from 98.6 to 101.4\,kPa over two months of experiments, so a fixed value would bias $A_n$. A sealed, rigid chamber may also build up pressure during a \textsc{read} as its air warms and the leaf adds water vapor. To decide whether each chamber needs its own barometer, we placed one air pressure sensor BMP280 inside an operating chamber and another outside for two days. Both aligned with the pressure records from two nearby airport stations (Fig.~\ref{fig:pressure}a). Across 745 closures, the chamber stayed within a few pascals of the room (mean $-1.7$\,Pa, 90\% of closures between $-5.6$ and $+2.4$\,Pa, Fig.~\ref{fig:pressure}b), less than 0.1~ppm (the S88 reading changes by 1.6\% per kPa)~\cite{senseair_s88_integration}. We therefore omit the in-chamber barometer and use the room pressure at each closure or a nearby weather station's reading adjusted to the lab's elevation.

\subsection{Adaptable Chamber and Leaf Clamps Fabrication}
\label{sec:geometry}

\begin{figure}[H]
  \centering
  \includegraphics[width=\linewidth]{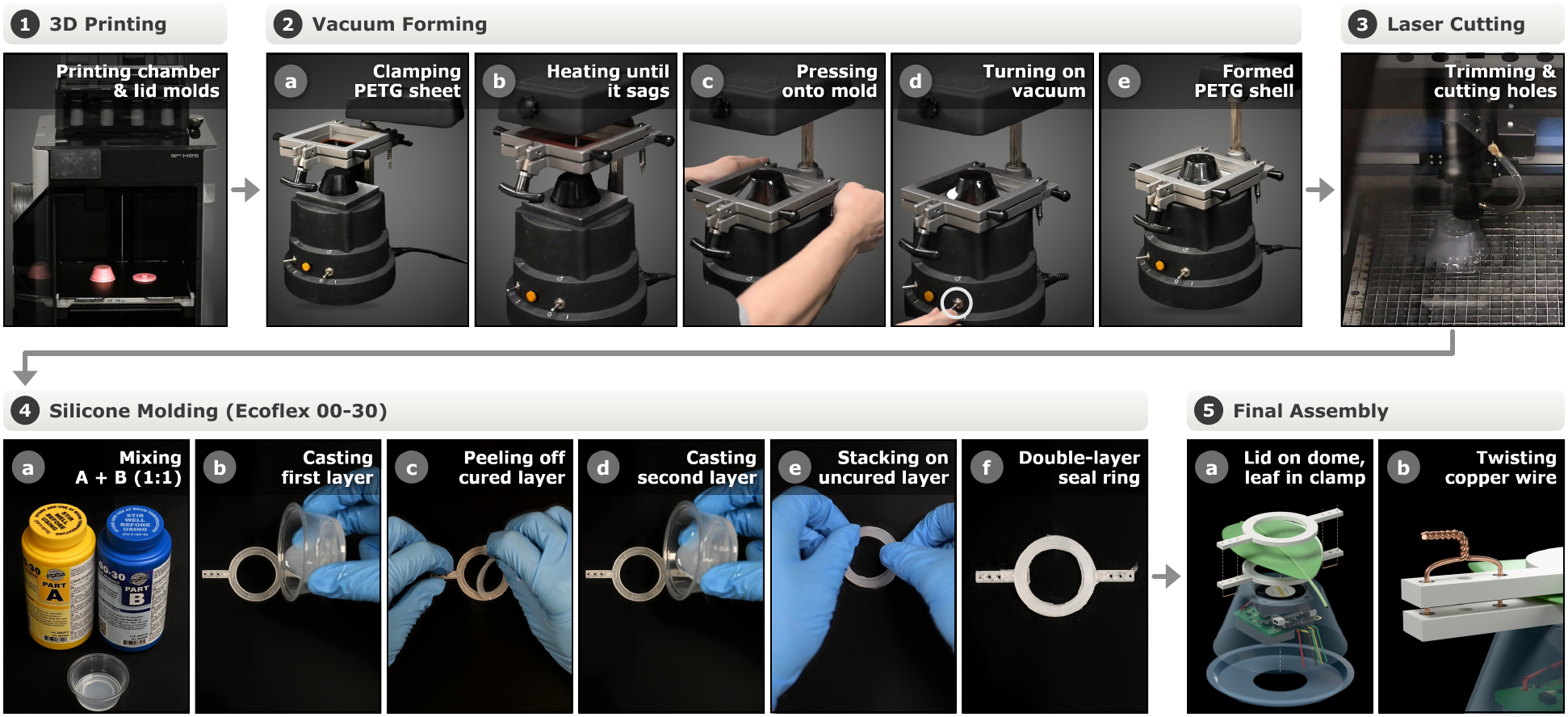}  
  \caption{Chamber fabrication: vacuum forming the PETG dome over a 3D-printed mold, laser cutting, and two-layer silicone casting of the seal ring (\ref{sec:geometry}), all other structural parts are printed and assembled directly.}
  \Description{}
  \label{fig:fabrication}
\end{figure}

\paragraph{Chamber Thermoforming.} We thermoform the chamber rather than print it. A thin-walled fused-filament
print leaks through gaps between its extrusion lines and
layers~\cite{szczech2024leakage}, and its layer lines scatter light, whereas a
formed PETG sheet is continuous, airtight, and clear enough to let light reach a leaf
enclosed in the dome (\S\ref{sec:app-adapt}). The printer thus makes only the
mold, which needs to be neither airtight nor clear, and a new leaf shape needs
only a new mold. 
The transparent chamber comprises an upper dome and a lower base joined by a snap-fit rim (Fig.~\ref{fig:hardware}c).
We vacuum-form the dome from a 1~mm rigid PETG sheet over a two-part 3D-printed mold using a desktop vacuum former.
The PETG sheet is clamped into the forming frame, with the mold centered on the vacuum-forming table. To ensure even heating, the heating element is preheated away from the sheet and then moved above it.
Once the softened sheet sags approximately 1~cm (Fig.~\ref{fig:fabrication}.2b), the frame is lowered over the mold and the vacuum pump engaged. The heater remains overhead for 5~s before retraction.
After the sheet conforms to the mold, we turn off the pump and allow the PETG to cool in place to prevent warping during demolding.
Because repeated thermal cycling gradually deforms the 3D-printed mold, we replace the mold after approximately 10 forming cycles.
The remaining structural parts, including the support frame, rack and gear, servo mount, and clamp plates, are 3D-printed in PLA and assembled without further processing. Design files and firmware will be released.

\paragraph{Clamp Aperture and Geometry.}
\label{sec:clamp}
The default leaf clamp uses a 30~mm circular aperture, enclosing $S = 7.07$~cm$^2$ of leaf area (Fig.~\ref{fig:hardware}c). 
This aperture size was informed by commercial leaf cuvettes, such as the $2 \times 3$~cm ($6.0$~cm$^2$) LI-COR LI-6800 and the $4.5$~cm$^2$ CIRAS-4 cuvette.
We selected a comparable scale to accommodate a range of leaf sizes while retaining a large enclosed area. 
For a fixed chamber volume $V$, a larger leaf area $S$ produces a faster CO$_2$ change ($-dC/dt$) for a given $A_n$ (Eq.~\ref{eq:simple}), improving resolution when assimilation is low.
To support narrower or smaller leaves, LeafBreath provides interchangeable clamp plates with smaller circular (23.9 mm diameter, 4.50~cm$^2$) and rectangular (25 × 18 mm, 4.50~cm$^2$) apertures, and a larger circular aperture (40 mm, 12.6~cm$^2$) for broad leaves (Fig.~\ref{fig:hardware}c). The chamber has a circular footprint, following the circular clamp used in the initial design. The same chamber body accommodates interchangeable clamp plates with circular or rectangular apertures, allowing the measurement area and aperture shape to be adapted to different leaves.

\paragraph{Leaf Clamp and Gasket Assembly.}
The clamp plates are 3D-printed in PLA with a circular groove around the aperture rim for a silicone gasket and two fastening tabs (Fig.~\ref{fig:hardware}b, c).
To form a thick and compliant silicone gasket, we use a two-step casting method: we first cast a silicone layer in the groove (Fig.~\ref{fig:fabrication}b) and remove it, then refill the groove with fresh silicone and press the cured layer back before it sets. This bonds the layers into a thickened gasket that ensures an airtight seal against the leaf surface.
Each tab has three holes, counted from the outer edge inward. During deployment, 24-gauge copper wires pass through the first and third holes and are twisted to secure the clamp; for larger leaves reaching the inner hole, the first and second holes are used instead.

\subsection{Airflow and Mixing}
\label{sec:airflow}
Mixing air in a closed chamber is necessary for two reasons. First, the leaf consumes CO$_2$ from the air adjacent to its surface. In still air, a CO$_2$-depleted boundary layer forms near the leaf, limiting replenishment by diffusion and potentially restricting photosynthesis. Second, the CO$_2$ sensor reads only nearby air. The flux calculation assumes this reading represents the whole chamber, which requires mixing.
A 30 mm axial fan runs continuously inside each chamber. It breaks the boundary layer at the leaf, mixes air during \textsc{read}, and speeds exchange with room air during \textsc{rest}. The fan is glued at mid-height, oriented upward toward the leaf.
Early prototypes used a 25 mm, 5 V WINSINN 2510 fan. With the chamber closed for 10\,min and no external light source, it warmed the chamber air by a median of 2.4\,$^\circ$C. We therefore replaced it with a 30 mm, 5 V Easycargo 3007 fan. Under the same conditions, the new fan warmed the air by 0.3 to 0.5\,$^\circ$C. Minimizing fan-induced warming is critical because temperature changes during \textsc{read} can cause incorrect slopes due to CO$_2$ sensor temperature drift (\S~\ref{sec:char-temp}).

\subsection{Chamber Sealing and Moisture Control}
\label{sec:sealing}
Each chamber requires two reliable seals: one at the top leaf clamp and another at the bottom movable lid. At the top aperture, the clamp rings with silicone gaskets press gently against the leaf to form a seal around the enclosed area. At the bottom, the chamber base seals against a servo-driven silicone lid during each \textsc{read}. 
Compared to the top gasket, the bottom base exposes the chamber interior to a much larger silicone surface. This silicone surface can absorb and desorb water vapor, changing internal humidity and gas measurements. We cover this lid with polyethylene cling wrap to limit moisture release and the resulting temperature variation, which can bias the CO$_2$ readings (\S~\ref{sec:char-temp}). Finally, to prevent leaks along the sensor and fan wiring, we solder thin individual copper wires directly to the component pins, route them flat through the snap-fit rim without overlapping, and seal the cable pass-throughs with hot glue. We characterize the overall leak rate $k$ with a real leaf clamped in place to simulate the real measurement conditions (\S~\ref{sec:char-leak}).

\subsection{READ--REST Cycle}
\label{sec:readrest}
The \textsc{read} duration balances the need for sufficient data to estimate the CO$_2$ slope against the changes in chamber conditions caused by the enclosure. We account for two different scenarios to determine the duration of \textsc{read}: the chamber reopens after three readings fall at least 100~ppm below the first reading of the \textsc{read}, or after 2~min, whichever occurs first. Faster photosynthesis therefore leads to shorter closures. The threshold applies only to drawdown. Measurements with rising CO$_2$, such as during net respiratory release, continue until the timeout.
During \textsc{rest}, the bottom lid opens to allow air exchange between the chamber and its surroundings. The lid remains open for 2 minutes, before the next closure. 

\subsection{Control Electronics}
\label{sec:electronics}
We used either an Arduino Mega 2560 or an ESP32-WROOM-based development board as the LeafBreath system controller.
In the Arduino implementation, the CO$_2$ and temperature and humidity sensor modules are powered through the board's 5~V pin. The board receives external power through its DC barrel jack. A separate DC-DC buck converter draws power from the VIN rail and supplies 5~V to the servo and fan. A resistive voltage divider reduces the Mega's 5~V TX signal to approximately 3.3~V for the S88's RX input, while the sensor's TX connects directly to the Mega's RX pin.
In the ESP32 implementation, the sensors, servo, and fan share a 5~V supply connected to the development board's 5~V pin. The power and ground conductors of a USB cable are soldered to the board's 5~V and GND pins, respectively, allowing the system to operate from either a USB wall adapter or a USB power bank.
Data were then read from the serial monitor or written to SD card.

\section{LeafBreath Characterization}
\label{sec:char}

\subsection{Chamber Volume Characterization}
\label{sec:char-vol}
After thermoforming each chamber, we laser-cut the leaf-clamp opening (Fig.~\ref{fig:fabrication}.3) and used hot glue to attach the clamp to the top of the chamber. We then measured the gross volume by filling the chamber with water and weighing the water. The chamber base is laser-cut after the volume measurement. We subtracted 8.75\,mL for the internal components to obtain the free gas volume. Gross volumes ranged from 86.2 to 93.5\,mL across the chambers. Each chamber's flux was calculated using its own measured volume.

The volume of the internal hardware (S88 GH CO$_2$ sensor, SHT40 temperature and humidity sensor, fan, and hot glue blob used to secure the fan to the chamber) was determined using water displacement. We submerged the components in a known volume of water to measure volume difference by water displacement. The displaced water was drawn into a graduated glass pipette using a rubber suction bulb. Each measurement was repeated three times.

\subsection{Chamber Leakage Characterization}
\label{sec:char-leak}
\begin{figure}[H]
  \centering
  \includegraphics[width=\linewidth]{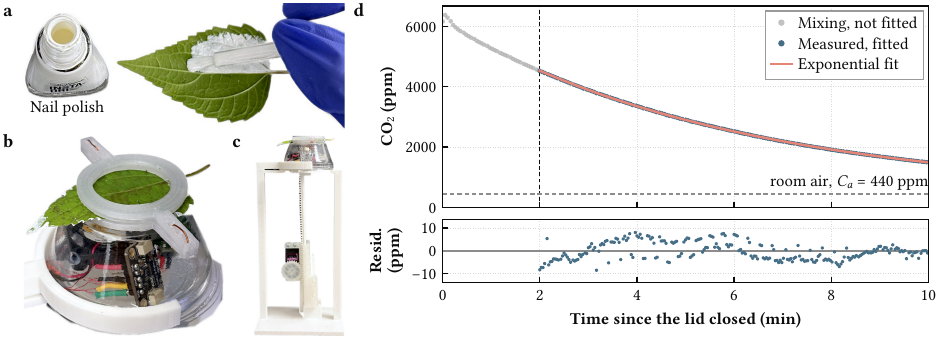}
  \caption{Leak measurement (\S~\ref{sec:char-leak}), one closure of one chamber with a real leaf clamped in place. (a) Apply nail polish to the bottom side of the leaf. (b) Clamp the leaf to the LeafBreath chamber, the bottom side facing the chamber. (c) The servo seals the chamber. (d) Top: after exhaled air raises the chamber CO$_2$, the decay toward room air is fitted with a single exponential starting 2 min after closure (earlier samples, during mixing, are excluded), for this closure $k = 2.82\times10^{-3}$ s$^{-1}$ ($\tau$ = 5.9 min). Bottom: fit residuals (RMSE 3.7 ppm).}
  \Description{}
  \label{fig:leak}
\end{figure}

As noted in \S~\ref{sec:sealing}, potential leak paths include the leaf clamp, lid seal, and snap-fit joints. We tested leakage with a real leaf to include the effects of leaf thickness and veins on the seal. We coated its lower surface with nail polish to suppress stomatal CO$_2$ exchange~\cite{wu2017using}, so that any change in chamber CO$_2$ comes from leakage alone. \emph{The nail polish is used only in this leak test; leaves measured with LeafBreath are never coated.}
For each test, we exhaled into the open chamber to raise CO$_2$ to 2000--4500\,ppm. We then sealed the chamber with the servo-driven rack for 10\,minutes with the fan running and recorded the decay toward room CO$_2$. Assuming $A_n=0$, Eq.~\ref{eq:flux} reduces to $dC/dt=k(C_a-C)$, with the solution
\begin{equation}
  C(t) = C_a + (C_0 - C_a)\,e^{-kt}
  \label{eq:leak}
\end{equation}
(Fig.~\ref{fig:leak}d) shows an example of the leakage curve. We fitted $k$ and $C_0$ for each closure. We fixed $C_a$ at 440\,ppm, the room CO$_2$ read by the chamber sensors before the breath was added (420 to 470\,ppm). Changing $C_a$ by 30\,ppm changed $k$ by at most 3\%.
We excluded the first 120\,s to allow the exhaled air to mix. Each chamber underwent three closures.
Single-exponential fits described the measured decay with fit residuals RMSEs of 1.3--4.7\,ppm. Across chambers, $k$ ranged from 0.41 to $2.80\times10^{-3}$\,s$^{-1}$, corresponding to time constants of 6--41\,min. This variation is consistent with differences in assembly and sealing.

\subsection{CO$_2$ Sensor Gain and Response Time}
\label{sec:char-co2}
LeafBreath estimates $A_n$ from the rate of CO$_2$ change, $dC/dt$, measured by a single sensor throughout each \textsc{read}. Sensor offset and gain errors vary across devices. A constant offset cancels out when calculating the slope, whereas a gain error directly scales the measured slope. In addition to these calibration errors, the sensor reading takes time to catch up with the changing chamber concentration. Long sensor response time causes deviations between the actual rate of CO$_2$ change and the measured slope during short \textsc{read}s. We used controlled CO$_2$ sweeps to characterize sensor gain and assess whether the sensors responded quickly enough to accurately capture CO$_2$ changes of approximately 100\,ppm, which equals the drawdown limit of the LeafBreath chamber closure (\S~\ref{sec:readrest}).

\begin{figure}[H]
  \centering
  \includegraphics[width=\linewidth]{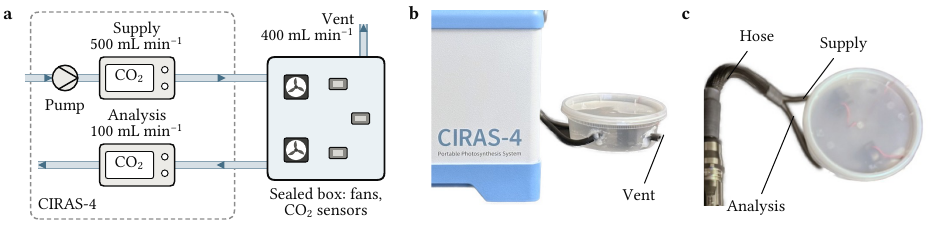}
  \caption{(a) The CIRAS-4 supplies controlled CO$_2$ to a fan-mixed container holding the CO$_2$ sensors and samples its air through a second line, excess flow leaves through a vent. (b) Front view of the setup.  (c) Top view, showing the supply and analysis lines. }
  \Description{}
  \label{fig:sweepsetup}
\end{figure}

We used the CIRAS-4 Leaf Cuvette Simulator (Fig.~\ref{fig:sweepsetup}c), a tool for the CIRAS-4 system for troubleshooting leakage. As shown in Fig.~\ref{fig:sweepsetup}a, we connected the supply air tube and analysis air tube to a custom 8-US-fl-oz (approximately 237\,mL) container. This arrangement allowed the CIRAS-4 to supply controlled CO$_2$ and measure the CO$_2$ in the container. The air supply delivered 500\,mL\,min$^{-1}$, while the analysis channel sampled container air at 100\,mL\,min$^{-1}$. A vent made up of a viton tube with 1\,cm long, 3/16\,in (4.8\,mm) inner diameter allows the remaining 400 \,mL\,min$^{-1}$ to escape the chamber. CO$_2$ sensors were spaced apart inside the fan-mixed container.

Before testing, we supplied air with 400\,ppm CO$_2$ to the container and calibrated each sensor at this CO$_2$ level using the manufacturer's software. We then stepped the supplied CO$_2$ upward and downward over a range of 0--2000\,ppm, including 100\,ppm increments per unit time from 0\,ppm to 1000\,ppm and a single increment from 1000 to 2000\,ppm. Throughout the sweeps, we recorded the CIRAS-4 reference-channel readings for the supplied air, the analysis-channel readings for the container air, and the readings from each S88 GH CO$_2$ sensor.

\begin{figure}[H]
  \centering
  \includegraphics[width=\linewidth]{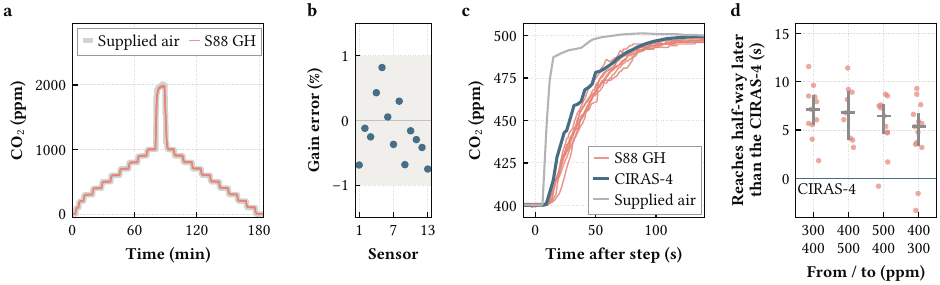}
  \caption{CO$_2$ sensor gain and response time (\S~ \ref{sec:char-co2}), measured against the CIRAS-4 in a fan-mixed container. (a) One staircase of supplied CO$_2$ steps and an S88 GH reading. (b) Gain error of each of the 13 sensors, from settled sensor readings, shaded band: ±1\%. (c) Response to a 400→500 ppm step: S88 GH sensors, the CIRAS-4 analysis channel, and the supplied air. (d) Delay of each S88 GH relative to the CIRAS-4 in reaching the step midpoint, for four transitions, dots: each sensor's median delay, black lines: median and inter-quartile range across sensors.}
  \Description{}
  \label{fig:sweep}
\end{figure}

Using the settled S88 GH sensor readings between 100 and 1000\,ppm, we estimate sensor offset and gain. Fitted gains were within ±1\% across the 13 sensors (Fig.~\ref{fig:sweep}b). To characterize response speed, we compared the sensor readings with the CIRAS-4 analysis-channel readings during the CO$_2$ transitions (Fig.~\ref{fig:sweep}c).

The S88 GH sensors measured air directly inside the container, whereas the CIRAS-4 analysis channel drew in container air through a sampling tube before measuring. This sampling path can introduce transport delay, which explains why some S88 responses appeared earlier than the CIRAS-4 measurements.

We assessed response speed during the 300$\leftrightarrow$400 and 400$\leftrightarrow$500\,ppm transitions (Fig.~\ref{fig:sweep}c, d). Across the four transition directions, the median sensor-level midpoint differences ranged from 5.4 to 7.2\,s, with some S88 responses reaching the midpoint earlier than the CIRAS-4 analysis channel.

Given these observations, we include the data recorded before 30 s when calculating $A_n$, rather than discarding the first 30\,s solely on the basis of the specified $T_{63}<30$\,s.

\subsection{Temperature-Induced Sensor Drift}
\label{sec:char-temp}
During each \textsc{read}, the temperature inside a chamber may change due to heat generated by both the hardware and the light source of plants. These temperature changes introduce a bias in the sensor readings, causing the measured CO$_2$ slope to deviate from the actual rate of change. 
We use the same container in \S~\ref{sec:char-co2}, supply constant 400\,ppm CO$_2$. By changing the air temperature in the container, we observed the effect of temperature on S88 GH sensor readings. We therefore evaluated its impact on $A_n$ estimation. 

The container included two mixing fans, an S88 GH CO$_2$ sensor, an SHT40 temperature and humidity sensor, and a BMP280 air pressure sensor. A hot plate was used to heat the container uniformly from the bottom of the container. We heated the container from 20\,$^\circ$C to 33\,$^\circ$C, maintaining a constant heating rate of approximately 0.41\,$^\circ$C/min to allow the S88 GH sensor to respond. The analysis-channel readings for the container air and the readings from sensors were recorded during the temperature sweeps.

\begin{figure}[H]
  \centering
  \includegraphics[width=\linewidth]{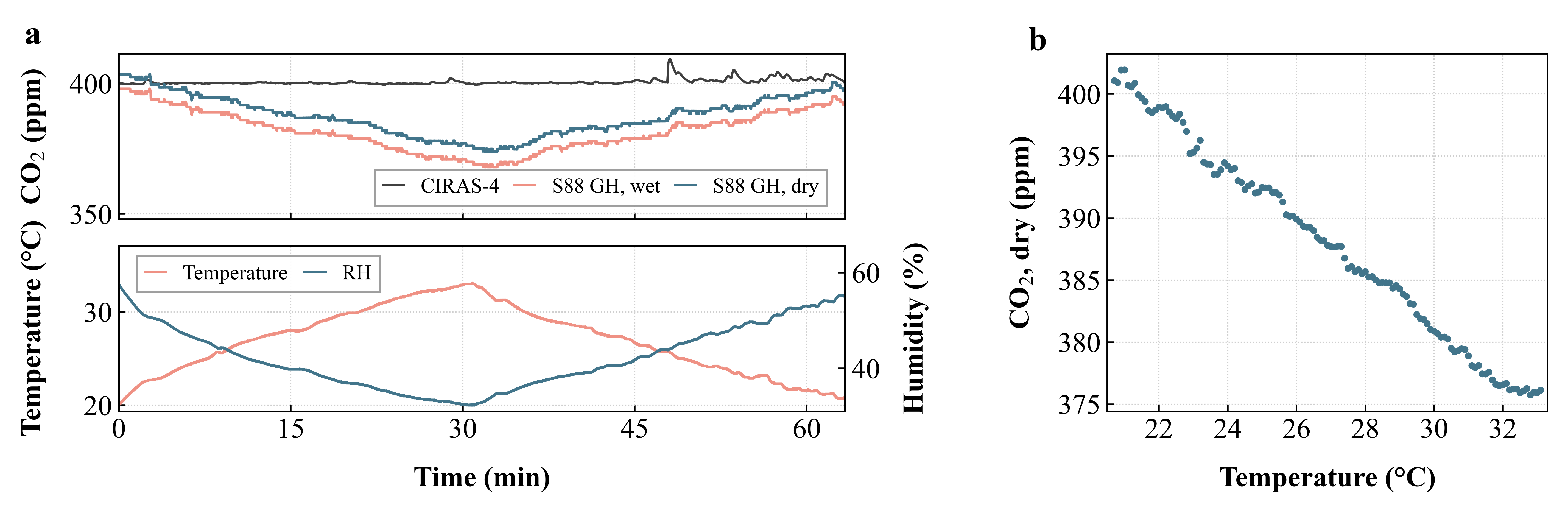}
  \caption{Temperature dependence of S88 GH CO$_2$ sensor readings. (a) Time series of S88 GH, SHT40, and CIRAS-4 analysis channel readings. Both wet-air and dry-air CO$_2$ readings are shown. (b) Dry-air CO$_2$ readings as a function of chamber temperature.}
  \Description{}
  \label{fig:Tsweep}
\end{figure}

Fig.~\ref{fig:Tsweep} shows the CO$_2$ readings and environmental conditions during the sweep. Pressure remained stable under temperature variation, ranging from 100.39 to 100.45\,kPa. Similarly, the wet-air and dry-air CO$_2$ readings exhibited a nearly constant difference (Fig.~\ref{fig:Tsweep}a). These observations suggested that changes in water vapor and air pressure had only limited effects on the S88 reading drift.

The S88 GH sensor readings showed a nearly linear temperature dependence, with its CO$_2$ readings decreasing by approximately 2.45\,ppm for every 1\,$^\circ$C increase between 22 and 25\,$^\circ$C (Fig.~\ref{fig:Tsweep}b). Based on the 29 sessions of data from the A–B–A test (\S~\ref{sec:aba}), the median rate of temperature increase was 0.38\,$^\circ$C/min. When calculating $A_n$, the change in the derivative of CO$_2$ readings at 20\,s decides the bias from the CO$_2$ sensor drift (\S~\ref{sec:flux-esti}). 
Assuming a 2.45\,ppm/$^\circ$C temperature sensitivity and a 0.38\,$^\circ$C/min rate of temperature change, the CO$_2$ sensor drift would contribute 0.0155\,ppm/s to the measured CO$_2$ slope, which is negligible to our $A_n$ estimation.

\subsection{CO$_2$ Slope Estimation and Flux Calculation}

\begin{figure}[H]
  \centering
  \includegraphics[width=\linewidth]{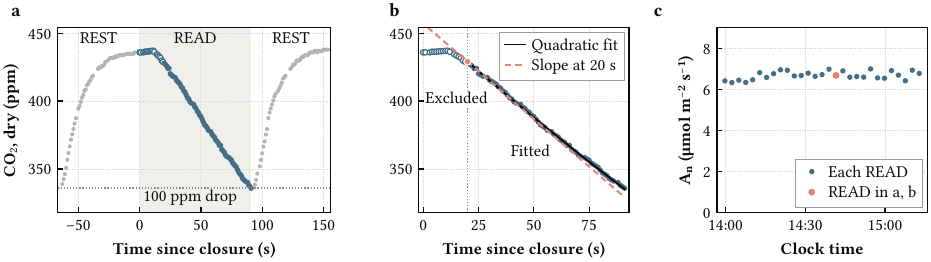}
  \caption{(a) Chamber CO$_2$ over a \textsc{rest}--\textsc{read}--\textsc{rest} cycle. The leaf draws CO$_2$ down by 100\,ppm during the closure. (b) The first 20\,s are excluded, a quadratic is fitted to the rest, and its slope at 20\,s yields $A_n$. (c) $A_n$ from consecutive \textsc{reads}, the highlighted point is the \textsc{read} in (a) and (b).}
  \Description{}
  \label{fig:read_cycle}
\end{figure}

Each $A_n$ value corresponds to one \textsc{read}, and is calculated using the system's individually characterized gain, volume and leak rate. We first apply the sensor's gain correction (\S~\ref{sec:char-co2}) and convert the CO$_2$ readings to a dry-air basis (Eq.~\ref{eq:flux}). We then fit the corrected CO$_2$ \textsc{read} readings to a quadratic curve from 20\,s onward and evaluated its derivative at 20\,s (Fig.~\ref{fig:read_cycle} b). Readings before 20\,s are excluded to allow the air inside the chamber to mix and the CO$_2$ sensor to respond. Although a slope at the edge of the fit is less stable than one at its center, use the slope at 20\,s gives the slope near the start of chamber closure, when the chamber CO$_2$ is still close to room levels (median 8\,ppm below the initial \textsc{read} value). 
This slope enters Eq.~\ref{eq:flux}, with the chamber's measured volume (\S~\ref{sec:char-vol}) and leak rate (\S~\ref{sec:char-leak}) to decide $A_n$. Ambient CO$_2$ value, $C_a$, is taken from the readings at the chamber closure (\S~\ref{sec:theory}).

\label{sec:flux-esti}

\section{Validation}
\label{sec:validation}
\subsection{Validation Against the CIRAS-4 Reference Instrument}
\label{sec:aba}

\begin{figure}[H]
  \centering
  \includegraphics[width=\linewidth]{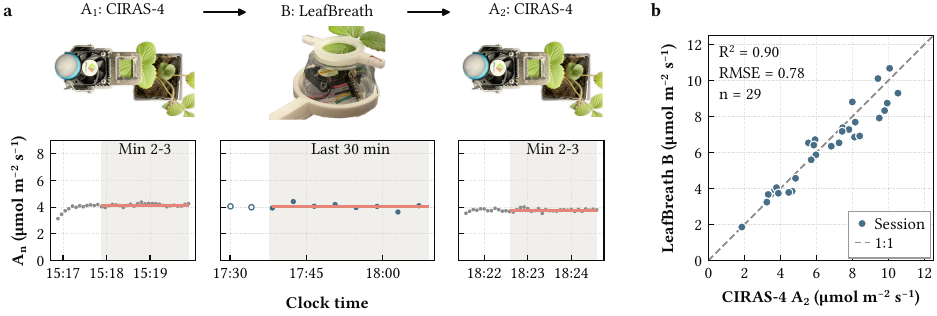}
 
  \caption{A–B–A validation (\S~\ref{sec:aba}). Each point is one session: (b) LeafBreath B against A$_2$ for all 29 sessions. Dashed line: 1:1. RMSE is the root-mean-square difference between B and A$_2$.}
  \Description{}
  \label{fig:aba}
\end{figure}

We validated LeafBreath against a CIRAS-4 portable gas-exchange system (PP Systems) using an A-B-A design. Each leaf was measured on CIRAS-4 (A$_1$), then with LeafBreath (B), then on CIRAS-4 again (A$_2$). Each leaf thus served as its own reference, while A$_1$ and A$_2$ showed whether it changed during the session. We ran 29 sessions over 12 days, testing 20 leaves from 7 strawberry plants. Five LeafBreath systems were used, all with 30\,mm clamps.

A fair comparison also required the leaf to receive the same light in both instruments. Under the grow lamp, the CIRAS-4 leaf sits beneath a clear cuvette top, while LeafBreath leaves its upper surface uncovered. We therefore used the CIRAS-4's outside sensor to measure photosynthetically active radiation (PAR) at each LeafBreath location. For each CIRAS-4 measurement, we moved the cuvette and adjusted its height to match the light at that location. The inside sensor had to read within 20\,$\mu$mol\,m$^{-2}$\,s$^{-1}$ of the outside reading. PAR inside the cuvette ranged from 280 to 560\,$\mu$mol\,m$^{-2}$\,s$^{-1}$ (median 430).

Room sensors logged CO$_2$, temperature, and humidity, while an infrared thermometer measured leaf temperature. We used these values to set CIRAS-4 to the conditions the leaf experienced in the room. Across readings, reference CO$_2$ ranged from 440 to 500\,ppm and leaf temperature from 21.6 to 25.8\,$^\circ$C. CIRAS-4 values were recorded every 5\,s for 2 to 3 minutes, with A$_2$ taken as the final-minute mean. B was the mean of LeafBreath \textsc{reads} in the last 30\,min before the leaf was moved to CIRAS-4. A$_2$ was taken within 20\,min of the last \textsc{read}.

Over an $A_n$ range of 1.9 to 10.5\,$\mu$mol\,m$^{-2}$\,s$^{-1}$, B agreed with A$_2$ (Fig.~\ref{fig:aba}). The comparison yielded $R^2 = 0.90$ and RMSE of 0.78\,$\mu$mol\,m$^{-2}$\,s$^{-1}$. The mean B/A$_2$ ratio was 0.97 (SD 0.10, 95\% CI 0.93--1.01). In the 24 sessions with A$_1$, the two CIRAS-4 readings were a median of 1.7\,h apart. A$_2$/A$_1$ ranged from 0.75 to 1.40, with no overall trend (geometric mean 1.00, 95\% CI 0.94--1.06). Individual leaves changed by a median of 9\%, which is why we compared B with A$_2$, the reading closest in time. B/A$_2$ also varied less than B/A$_1$ (SD 0.10 vs 0.17). Sessions are not fully independent: the 29 sessions span 20 leaves from 7 plants, with 7 leaves measured on more than one day. Using only the first session of each leaf gave similar agreement ($R^2 = 0.92$, RMSE 0.74\,$\mu$mol\,m$^{-2}$\,s$^{-1}$, $n = 20$).

\subsection{Biological Validation using C3 and CAM Plants}
\label{sec:c3cam-validation}
\begin{figure}[H]
  \centering
  \includegraphics[width=\linewidth]{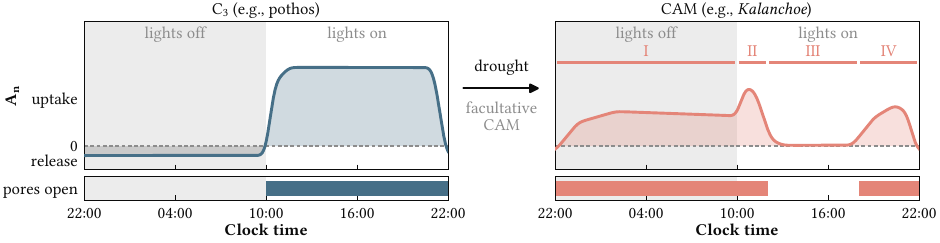}
  \caption{How C$_3$ and CAM leaves look in $A_n$ over one day (schematic), on the clock of our experiments (lights on 10:00 to 22:00, gray: lights off), bars below: when the leaf pores are open. A C$_3$ leaf (left) releases a little CO$_2$ in the dark and takes up CO$_2$ only in the light, when its pores are open. A CAM leaf (right) opens its pores at night and takes up CO$_2$ in the dark (I), takes up a short burst when the light comes on (II), keeps its pores shut at midday while it uses the CO$_2$ stored overnight (III), and takes up CO$_2$ again in the late afternoon (IV). Plants with facultative CAM, such as \textit{Coleus amboinicus} (\S~\ref{sec:app-research}), move from the left pattern toward the right one under drought.}
  \Description{}
  \label{fig:c3cam}
\end{figure}

Leaves take up CO$_2$ through small pores on their surface. Most plants, including wheat, rice, and soybean, use C$_3$ photosynthesis (Fig.~\ref{fig:c3cam} left): they open their pores in daylight and fix CO$_2$ as soon as it enters. At night they only respire, so a C$_3$ leaf takes up CO$_2$ while the light is on, where its $A_n$ is positive, and releases a little in the dark, where its $A_n$ sits slightly below zero. Plants with crassulacean acid metabolism (CAM) photosynthesis (Fig.~\ref{fig:c3cam} right), such as cacti, \emph{Aloe vera}, and \emph{Kalanchoe}, save water by reversing this timing. They open their pores at night, when the air is cooler and more humid, and store the CO$_2$ as acid. In daylight they close their pores and release the stored CO$_2$ inside the leaf for photosynthesis~\cite{luttge2004ecophysiology}. A CAM leaf therefore has a non-zero $A_n$ in the dark. Its day follows four phases~\cite{luttge2004ecophysiology}: uptake through the night (I), a short burst of uptake when the light comes on (II), near-zero exchange at midday while the pores stay shut (III), and renewed uptake in the late afternoon once the stored acid runs out (IV). Some plants switch between the two: they use C$_3$ when well watered and shift toward CAM under drought~\cite{winter2008nature}, which \S~\ref{sec:app-research} follows in \emph{Coleus amboinicus}. Telling these patterns apart needs $A_n$ day and night, which is what LeafBreath records.

\begin{figure}[H]
  \centering
  \includegraphics[width=\linewidth]{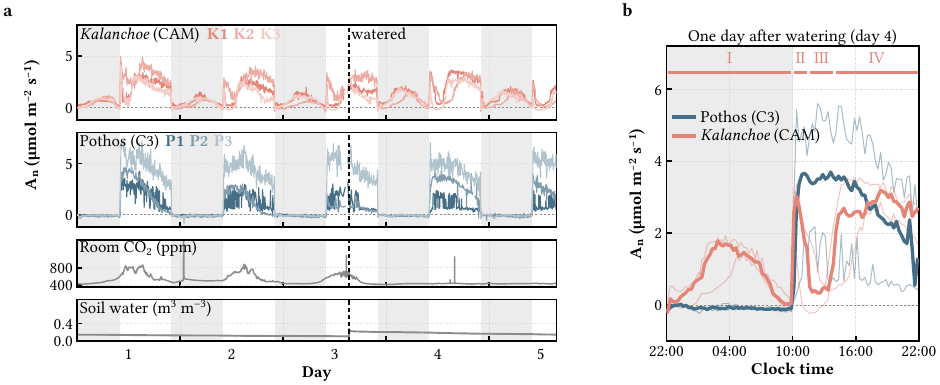}
  \caption{Five days of $A_n$ from two species in one room. (a) Three Kalanchoe (CAM) and three Pothos (C3) leaves, with room CO$_2$ and soil water content, gray: lights off, dashed line: watering. (b) Day 4, one day after watering: Kalanchoe takes up CO$_2$ at night and shuts down at midday, following the four CAM phases (I: nocturnal uptake, II: morning burst, III: midday stomatal closure, IV: afternoon reopening), while Pothos tracks the light, following the C3 pattern.}
  \Description{}
  \label{fig:kf_pothos}
\end{figure}
We then ran LeafBreath for five days on three \emph{Kalanchoe} and three pothos leaves in a classroom building (Fig.~\ref{fig:kf_pothos}). A grow light above ran from 10:00 to 22:00. Ambient CO$_2$, temperature, and humidity were not actively controlled, and normal indoor activities continued throughout the run. The chambers were left undisturbed, and the only visit was to water the pots on the evening of day 3.

On day 4, one day after watering, \emph{Kalanchoe}'s trace followed the four classic CAM phases: night uptake, a morning burst, midday closure, and afternoon reopening (Fig.~\ref{fig:kf_pothos}b). Pothos, which cannot store CO$_2$ overnight, only took up CO$_2$ while the light was on, so its $A_n$ is near zero after dark. LeafBreath captured the distinct day-night CO$_2$ exchange patterns associated with CAM and C3 photosynthesis side by side, under uncontrolled environmental conditions.

Since the room's environmental conditions were uncontrolled, the room CO$_2$ swung between about 400 and 800 ppm based on building occupancy. However, these swings have little effect on $A_n$ estimation. A swing moves where a reading starts, but since $A_n$ is determined from how fast the CO$_2$ drops during a closure, changing the starting point will not affect the calculation of this difference.
Our leak correction assumes constant room CO$_2$ during each closure. At $k \approx 10^{-3}$\,s$^{-1}$, a 1000\,ppm change can bias the slope by 1\,ppm\,s$^{-1}$, comparable to daytime leaf drawdown. We therefore reject closures if room CO$_2$ changes by more than 25\,ppm between 10\,s before closure and its end. Gradual changes over hours can be large overall but remain small within this short window.

\begin{figure}[H]
  \centering
  \includegraphics[width=\linewidth]{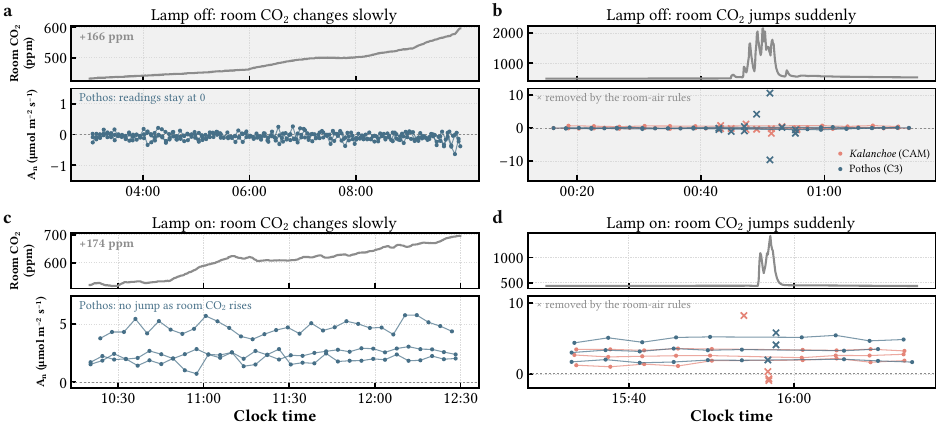}
  \caption{How room CO$_2$ affects the readings. Each panel shows room CO$_2$ (top) and $A_n$ (bottom) over the same period, gray background: light off. (a, c) Room CO$_2$ rising slowly by about 170\,ppm, light off (a) and on (c), only the Pothos leaves are shown. All readings carry on smoothly. (b, d) Brief spikes in room CO$_2$, with the light off (b) and on (d). Closures during the spikes ($\times$) are removed by the 25\,ppm criterion, those before and after are unchanged.}
  \Description{}
  \label{fig:room_co2}
\end{figure}

Fig.~\ref{fig:room_co2} shows both cases. When the room CO$_2$ rose slowly by about 170\,ppm over two hours (a) and seven hours (c), no closures were invalidated by this criterion. Readings carried on smoothly: near zero in the dark (a) and without jumps in the light (c). When a person breathed about 30\,cm from the chambers, the CO$_2$ reading jumped to 2150\,ppm within minutes. The $A_n$ of the closures during the jump read between $-10$ and $+11$\,$\mu$mol\,m$^{-2}$\,s$^{-1}$ and were removed, while the readings before and after were unchanged (b, d).

\section{Application Demonstrations}
\label{sec:application}
We show LeafBreath in three settings: a research scenario that follows many leaves over days, a scenario in which the chamber design adapts to different leaf geometries, and an education scenario that demonstrates a light response curve measuring session.

\subsection{Facultative CAM Progression Across Leaves}

\textit{Coleus amboinicus} changes how it photosynthesizes as water runs out. When well watered, its leaves follow the C3 pathway and take up CO$_2$ in daylight, like most plants (Fig.~\ref{fig:c3cam}, C3). Under drought, they shift toward CAM and take up CO$_2$ at night (Fig.~\ref{fig:c3cam}, CAM), and the shift reverses after re-watering. 

Winter et al. showed this shift on \emph{a single mature leaf} with $A_n$. To compare leaves of different ages, they manually punched discs from young and mature leaves at dusk and dawn, and titrated the acid stored overnight~\cite{winter2021constitutive}. Titration is laborious and destroys the leaf. It provides one value per leaf but no trend during a day, so whether young and old leaves shift or recover at different times remained unknown. To answer this question, $A_n$ values from several living leaves must be recorded continuously for days. This would otherwise take one gas-exchange instrument per leaf. LeafBreath provides this without moving an instrument from leaf to leaf.

\begin{figure}[H]
  \centering
  \includegraphics[width=\linewidth]{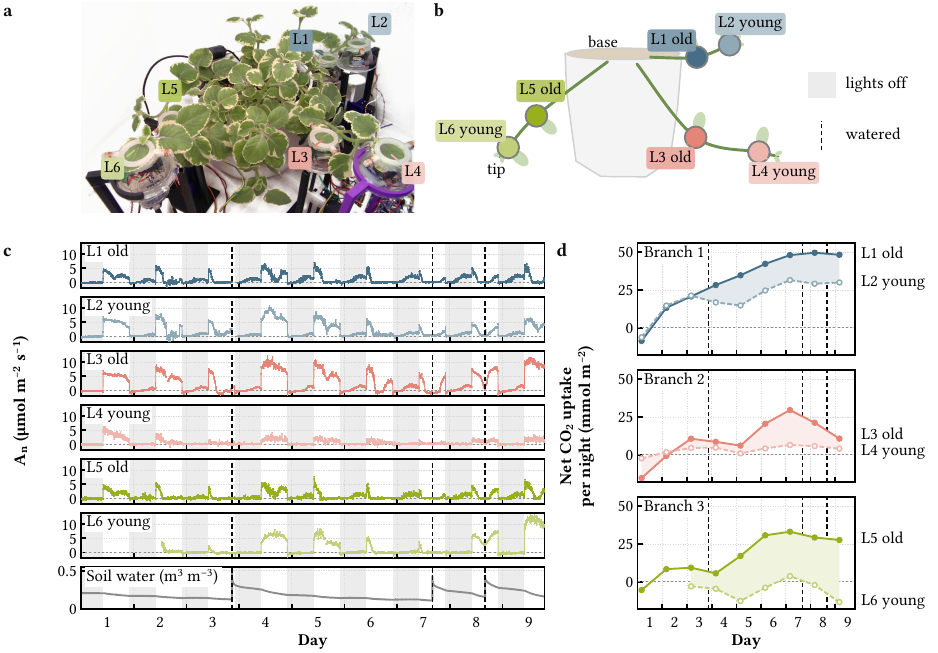}
  \caption{Nine-day deployment on six leaves of one Coleus amboinicus plant (\S~\ref{sec:app-research}). (a) The plant with six chambers. (b) Leaf positions: an older and a younger leaf on each of three branches, with node positions counted from the branch tip. (c) $A_n$ per leaf and soil water content, gray: lights off, dashed lines: watering. (d) Net CO$_2$ uptake per night for each leaf, grouped by branch. In (d), L6 has no data before 12:40 on day 2, when its lid started to close.}
  \Description{}
  \label{fig:coleus}
\end{figure}

We clamped six LeafBreath chambers on six leaves of one potted plant: on each of three branches, an older leaf nearer the base and a younger leaf nearer the tip (L1 to L6, Fig.~\ref{fig:coleus}a, b). We let the soil dry and watered the pot on days 3, 7, and 8.

\textit{Drought moved every leaf toward CAM.} As the soil dried (water content 0.21 to 0.13\,m$^3$\,m$^{-3}$ by day 3), daytime $A_n$ fell by a median of 99\% (Fig.~\ref{fig:coleus}c). At night the leaves did the opposite: on night 1 every measured leaf released CO$_2$, but by night 3 five of them took it up in the dark (Fig.~\ref{fig:coleus}d), which is the signature of CAM.

\label{sec:app-research}

\textit{Older leaves shifted further.} On every branch, the older leaf took up more CO$_2$ at night than the younger one, for example 48 versus 32\,mmol\,m$^{-2}$ for L1 and L2 on night 7. Winter et al. found the same direction in the acid that mature and young leaves stored overnight, here it appears in the CO$_2$ taken up from the air, night by night.

\textit{Leaves recovered at different times.} Within a day of the first watering, every leaf took up CO$_2$ in daylight again, but night-time uptake did not reverse together: it kept rising in L1 and L5 for two more nights and fell in the other leaves. After the later waterings, L3 and L6 regained daytime uptake first, while L1, L2, and L5 stayed near zero by day.

Leaves of one plant, in the same soil, thus entered and left CAM at different times, a difference that a whole-plant chamber would average away.

\begin{figure}[H]
  \centering
  \includegraphics[width=\linewidth]{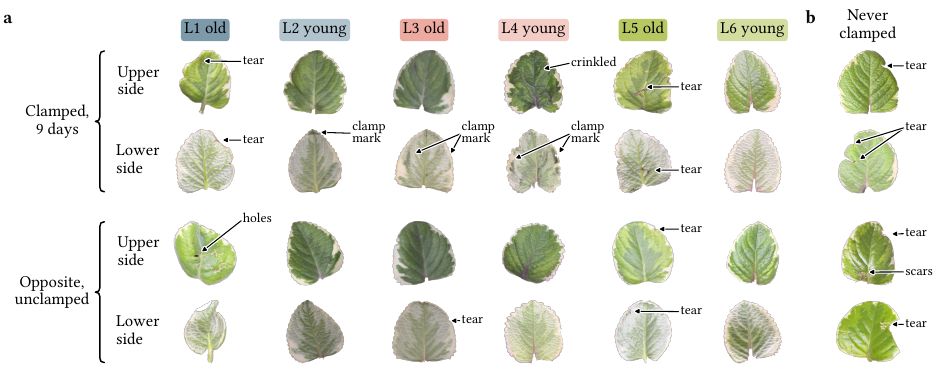}
  \caption{Coleus leaves after the nine-day run, cut out from their photographs. (a) Each clamped leaf (L1-L6) and the opposite leaf at the same node, which was never clamped, from the upper and the lower side. L1-L5 were photographed right after the run, L6 and its opposite leaf two days after. On the lower side of the opposite leaf of L1, a nail file used to hold the leaf flat covered part of it; the dashed line marks the estimated leaf edge. (b) Leaves of the same plant that were never clamped.}
  \Description{}
  \label{fig:coleus-clamped}
\end{figure}

LeafBreath is designed to support long-term data acquisition. To check how nine days of clamping affected the leaves, we photographed each clamped leaf and the opposite leaf, which had never been clamped (Fig.~\ref{fig:coleus-clamped}a). To drive the plant into CAM, the plant was exposed to stressful conditions: strong light intensity and withheld water for the first three days. The clamped leaves showed visible marks: small tears on L1 and L5, dark patches where the clamp pressed on the lower side of L2, L3, and L4, and a crinkled L4. Tears also appeared on the leaf opposite of L1 and on other leaves that were never clamped (Fig.~\ref{fig:coleus-clamped}b). This can be attributed to the brittle texture of the leaves that cause them to split when pressed or bent. The marks did not stop the leaves from working. On the last day, all six clamped leaves still took up CO$_2$ while exposed to light, and the older leaf of each pair still took up CO$_2$ at night (Fig.~\ref{fig:coleus}).

\subsection{Adapting LeafBreath to Different Leaf Geometries}
\label{sec:app-adapt}
\begin{figure}[H]
  \centering
  \includegraphics[width=\linewidth]{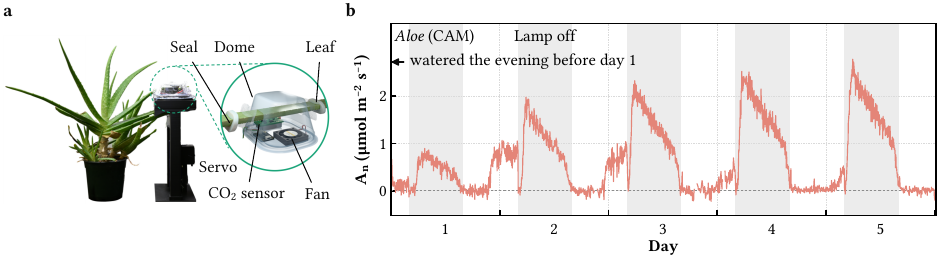}
  \caption{An \emph{Aloe vera} leaf measured for five days after watering. (a) The leaf passes through a custom dome, with silicone and cling wrap sealing the openings. (b) $A_n$ of every reading; grey, light off. Nighttime peaks increased from 0.8 to 2.6\,$\mu$mol\,m$^{-2}$\,s$^{-1}$.}
  \Description{}
  \label{fig:aloe}
\end{figure}

\begin{figure}[H]
  \centering
  \includegraphics[width=\linewidth]{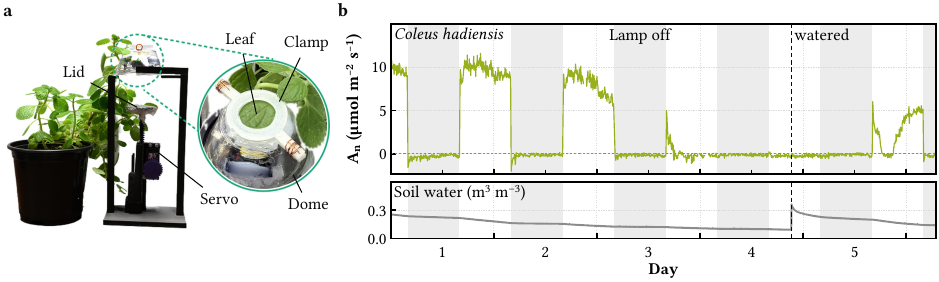}
  \caption{One \emph{Coleus hadiensis} leaf measured for five days. (a) Smaller clamp plates provide a 4.5\,cm$^2$ aperture that the leaf can cover completely. (b) $A_n$ of every reading, with soil water content from the same pot; dashed line, watering; grey, light off. Daytime uptake declined as the soil dried and returned after watering.}
  \Description{}
  \label{fig:hadiensis}
\end{figure}

We adapted the clamp plates and thermoformed dome to fit leaves of different sizes, shapes, and thicknesses (RQ1). Three variants held a thick \emph{Aloe vera} leaf passing through the chamber, a small \emph{Coleus hadiensis} leaf, and a whole basil leaf. The \emph{Coleus hadiensis} leaf used a 4.5\,cm$^2$ clamp, while the basil leaf was enclosed in the dome (Fig.~\ref{fig:aloe}--\ref{fig:hadiensis}). Each variant stayed in place for two to five days in the same indoor room, with the light on from 10:00 to 22:00.

\emph{Thick leaves with uneven shapes.} The \emph{Aloe vera} leaf was thick and did not fit the flat clamp. We formed a custom dome to fit its shape, with the leaf passing through the chamber (Fig.~\ref{fig:aloe}a). We pressed silicone and cling wrap into the holes around the leaf to seal the openings. The dome volume was 80.75\,mL, with 8.75\,mL occupied by internal hardware. We estimated the enclosed leaf volume as 4.5\,mL, giving a net air volume of approximately 67.5\,mL. The enclosed leaf area was also estimated, at approximately 12\,cm$^2$. Over five days, the record showed nighttime CO$_2$ uptake, with peaks increasing from 0.8 to 2.6\,$\mu$mol\,m$^{-2}$\,s$^{-1}$ after watering (Fig.~\ref{fig:aloe}b).

\emph{Small leaves.} The \emph{Coleus hadiensis} leaf was too small to cover the standard 30\,mm aperture. We replaced the clamp plates with a smaller design, reducing the aperture area to 4.5\,cm$^2$ (Fig.~\ref{fig:hadiensis}a). The smaller opening allowed the leaf to cover the aperture completely. We monitored one leaf for five days, recording a decline in daytime uptake as the soil dried and renewed uptake after watering (Fig.~\ref{fig:hadiensis}b).

These records also show that watering does not immediately restore photosynthesis. Nighttime uptake in \emph{Aloe vera} increased over several days after watering. In \emph{C.~hadiensis}, soil water content rose immediately, but uptake the next day remained below its initial level and followed a different daily pattern. Direct $A_n$ measurements capture these responses, which soil water measurements and the light schedule alone do not describe.

\begin{figure}[H]
  \centering
  \includegraphics[width=\linewidth]{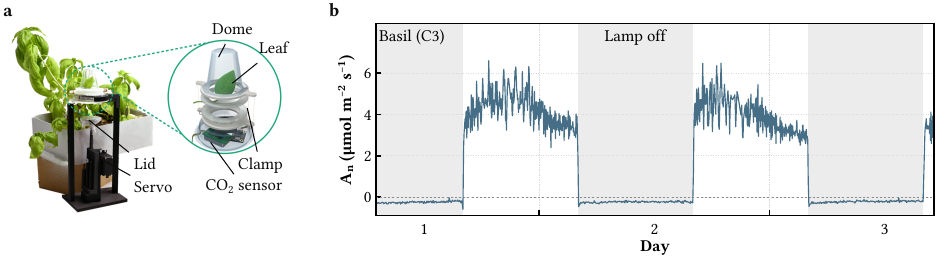}
  \caption{A whole basil leaf measured for two days. (a) A thermoformed cover encloses both leaf surfaces without flattening the leaf; the inset shows the chamber parts. (b) $A_n$ of every reading; grey, light off. The leaf took up CO$_2$ with the light on and released CO$_2$ at night.}
  \Description{}
  \label{fig:basil}
\end{figure}

\emph{Curved leaves.} For basil, we formed an upper cover that enclosed the whole leaf instead of pressing its curved surface into a flat clamp (Fig.~\ref{fig:basil}a). The leaf sat inside the chamber, allowing measurement of CO$_2$ exchange through both surfaces. Unlike the fixed clamp aperture, this design requires the leaf area to be measured separately. The record showed daytime CO$_2$ uptake and nighttime release over two days (Fig.~\ref{fig:basil}b).

\subsection{Educational Demonstration: Measuring Photosynthesis Across Light Levels}
\label{sec:app-education}

\begin{figure}[H]
  \centering
  \includegraphics[width=\linewidth]{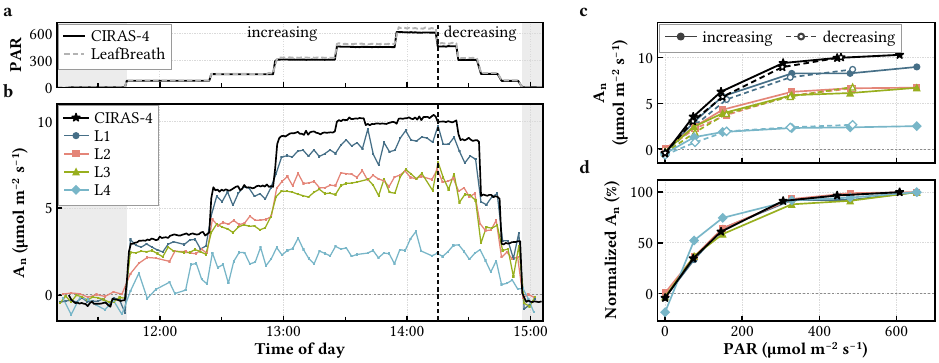}
  \caption{Light response curve measured with four LeafBreath chambers and a CIRAS-4, each holding a different strawberry leaf under the same light. (a) PAR at the CIRAS-4 and at the LeafBreath chambers (estimated from a shared ambient AS7341 light sensor). Light was raised from darkness in steps of 30\,min, then lowered in steps of 10\,min. PAR increased by 12.5\% for the first two steps and 25\% for the last three steps. (b) $A_n$ from every LeafBreath measurement cycle (colours) and from the CIRAS-4 (black, 1-min median). (c) $A_n$ at each light step, filled markers and solid lines, increasing light, open markers and dashed lines, decreasing light. (d) Normalized $A_n$ at each light step. Leaves are numbered by $A_n$ at the brightest step.}
  \Description{Four panels. (a) PAR over about four hours: dark, then steps of about 75, 150, 300, 450 and 600, then back down through 450, 300, 150 and 75 to dark. (b) Net CO$_2$ exchange over the same time: the CIRAS-4 leaf rises in steps to about 10, three LeafBreath leaves follow to about 8.5, 6.5 and 6, and a fourth stays near 2.5. (c) Net CO$_2$ exchange against PAR: all curves rise steeply up to about 300 and flatten above it, increasing and decreasing light give similar curves. (d) The same curves divided by each leaf's value at the brightest step nearly overlap, rising from about zero in the dark to one.}
  \label{fig:lightresp}
\end{figure}

A light response curve is a standard exercise in plant physiology courses that correlates light intensity with the value of $A_n$. We measured five strawberry leaves under a grow lamp, four in LeafBreath chambers and one in the CIRAS-4 (Fig.~\ref{fig:lightresp}). The light intensity was increased from darkness to the lamp's maximum level at intervals of  30\,min and then lowered in intervals of 10\,min for a total length of 4 hours.

Every leaf showed a similar trend: $A_n$ rose steeply before approaching saturation at a PAR of approximately 300\,$\mu$mol\,m$^{-2}$\,s$^{-1}$. For each leaf, we divided $A_n$ at each light level by that leaf's $A_n$ at the brightest step. The normalized curves overlapped (Fig.~\ref{fig:lightresp}d), except for L4 in the dark. This is because L4 had the lowest $A_n$, so dividing by it made its small dark value look large. In classrooms, instructors could demonstrate this experiment before student groups independently measure and compare light-response curves on their own leaves.

\section{Discussion}
\label{sec:discussion}

\subsection{Revisiting  Closed-Chamber Measurements on Leaf Scale}
Open-flow systems offer environmental control, but their pumps, gas conditioning, and analyzers add cost and complexity. LeafBreath revisits the closed-chamber method by automating the closure and reopening that manually operated leaf chambers require, enabling unattended measurements over multiple days. Open-flow systems with separate reference and sample analyzers require matching to correct differences between their readings. LeafBreath instead tracks concentration changes over time with the same sensor, so a constant offset cancels from the estimated slope. This reduces reliance on absolute concentration accuracy and supports the use of a low-cost NDIR sensor.

\subsection{Trade-offs of Brief Closure}
When the chamber closes, the leaf continues to take up CO$_2$, lowering the CO$_2$ concentration in the chamber. This decrease can reduce photosynthetic uptake, causing the measured rate to differ from that under ambient CO$_2$. LeafBreath estimates the slope 20\,s after the start of each \textsc{read} to estimate $A_n$ before CO$_2$ falls far below the outside level. It ends the closure after a 100\,ppm decrease in CO$_2$ to limit further depletion (\S~\ref{sec:readrest}).
A larger chamber would slow CO$_2$ depletion, but the smaller concentration change over the same interval would be harder for the sensor to detect. A longer \textsc{rest} would reduce the fraction of time the chamber remains closed, but yield fewer measurements. A CO$_2$ sensor with a shorter response time could shorten each closure, as illustrated by the specially developed sensor and manually operated chamber in MIC-100~\cite{tanaka2022mic}. We chose an off-the-shelf CO$_2$ sensor so that, apart from the thermoformed and 3D-printed chamber parts, all components can be purchased, allowing others to rebuild LeafBreath.
 
\subsection{Limitations}
LeafBreath does not yet provide reliable transpiration measurements. Transpiration quantifies water loss from a leaf and, together with $A_n$, indicates how much carbon the leaf gains per unit of water lost~\cite{busch2024guide}. Although the cling-wrap limits moisture release from the silicone lid and the associated temperature drift (\S~\ref{sec:sealing}), chamber materials can still absorb water vapor, making the measured transpiration rate lower than the actual rate. The associated water-vapor dilution of CO$_2$ is accounted for by the dry-air correction (\S~\ref{sec:theory}, Eq.~\ref{eq:flux}), which uses measured chamber temperature and humidity without requiring a transpiration estimate. Future work could characterize how much water the chamber materials absorb or use materials that absorb less water.

The standard clamp-based design encloses only the lower leaf surface, where most plant species have their stomata~\cite{muir2015making}. Some species have approximately half their stomata on each surface, so lower-surface measurements can underestimate whole-leaf $A_n$. The whole-leaf chamber demonstrated with basil encloses both surfaces (\S~\ref{sec:app-adapt}), but introduces additional trade-offs. An upper PETG cover can reduce light reaching the leaf, make the device more cumbersome to wear, and potentially compress the stem. Leaf area must be determined by tracing the leaf outline rather than using a fixed aperture, and adequate air mixing around both surfaces needs to be verified. The basil configuration was not validated against CIRAS-4 because the curved leaves could not fit its flat chamber without being flattened.
The strawberry leaves used for validation have their stomata mainly on the lower surface, making them suitable for our standard chamber and clamp design.

The validation covers one species, one chamber design (77.5--84.8\,mL net), and only the 30\,mm clamp. Besides, the comparison with the CIRAS-4 is itself limited by manual clamping of LeafBreath: the operator who moved the leaf between instruments changed the CO$_2$ around the chamber with their breath. Therefore, LeafBreath measurements required a waiting period to allow the ambient CO$_2$ level to stabilize. More generally, sensitivity to ambient CO$_2$ fluctuation can delay LeafBreath measurements when the environment is disturbed.

\subsection{Toward Unattended Deployment}
Running unattended does not remove the need for visits. In an outdoor pilot run, a strawberry leaf came out of its clamp at night, probably disturbed by an animal: during a midnight, the light sensor was covered for about 45\,s while the chamber CO$_2$ jumped from about 420 to 4200\,ppm, indicating a change from a breath. The chamber went on measuring without a leaf until the next visit about ten hours later. Leaves and their stalks also grow, so a chamber fixed in place moves relative to its leaf and has to be checked and readjusted about once a week. When a leaf is not attached, the CO$_2$ and humidity would stop changing with \textsc{read} and \textsc{rest}. A chamber could therefore compare each \textsc{read} with its recent record and send a warning when the leaf stops exchanging gas, so that visits are made when needed rather than on a fixed schedule.

\subsection{Future Work}
Two hardware changes would make LeafBreath smaller and more efficient for battery-powered operation. First, the servo and its rack-and-pinion stand could be replaced by a bistable actuator on the chamber itself, which holds the chamber open or closed without power and draws current only while switching. Without the stand, a light stake or clip could hold the chamber, so leaves that a fixed stand cannot reach could also be measured. Second, a smaller, low-power CO$_2$ sensor such as the Senseair S12 ($18 \times 15 \times 7$\,mm, about a third of the volume of earlier Senseair sensors~\cite{senseair_s12}) would shrink both the chamber and the hardware inside it. A smaller gas volume over the same leaf area results in a larger CO$_2$ drawdown for the same $A_n$ (Eq.~\ref{eq:flux}), and so a stronger signal from weakly photosynthesizing leaves. With both changes, the fan and the controller would account for most of the power use.

\subsection{ Potential Impact of LeafBreath on Plant Biology and Ubicomp}
\label{sec:impact}

\noindent \textbf{1) Shift Destructive Sampling to Continuous Multi-Leaf Tracking:} LeafBreath’s autonomous, multi-leaf, and affordable monitoring unlocks data that were impractical to collect in plant biology before. For example, continuously tracking six leaves simultaneously during a drought experiment (\S~\ref{sec:app-research}) provides coverage that a carried instrument cannot. Until now, comparing CAM across leaf ages has required laborious leaf collection at dusk and dawn followed by destructive tissue titration \cite{winter2021constitutive}. LeafBreath shows the CAM comparison across leaf ages \emph{cheaply, continuously, and non-destructively}. To our knowledge, this is the first continuous gas-exchange record comparing individual leaves of different ages at the same time
(Fig.~\ref{fig:coleus}). A continuous record also shows when each leaf changes, providing richer knowledge than spot tests or destructive tissue analyses. Furthermore, commercial gas-exchange instruments cost over USD 42,000, limiting access to well-funded institutions. LeafBreath could democratize this capability for smaller plant biology and agriculture labs.

\noindent \textbf{2) Unlocking Novel Plant Research Questions and Scientific Discovery:}
Longitudinal measurements could help researchers understand why leaves of the same plant behave differently. They could also reveal how stress develops over time and how recovery varies across leaves. By removing cost and manual constraints, LeafBreath empowers researchers to explore new plant biology questions.
\textit{\textbf{i) Linking photosynthetic and genomic data:}} Continuous $A_n$ monitoring reveals changes in a leaf's photosynthetic activity over time. Instead of harvesting tissue at arbitrary times of day, researchers could use changes in $A_n$ to identify sampling times. For example, they could collect tissue when a monitored leaf begins taking up CO$_2$ at night or resumes daytime uptake after watering. This would link the molecular sample to the leaf's recent photosynthetic activity.
\textit{\textbf{ii) Pairing $A_n$ with other leaf measurements:}} P02 measures conductance and fluorescence on five leaves per plant across up to 50 plants. However, they measure $A_n$ on about three leaves per plant across at most 20 to 30 plants, ``due to time constraints.'' LeafBreath chambers left on those leaves could provide an $A_n$ record alongside these measurements. Researchers could then compare each marker reading with photosynthetic activity before and after it on the same leaf.

\noindent \textbf{3) Democratizing Photosynthesis and Plant Biology Education:}
Photosynthesis is a foundational topic in biology, but commercial gas-exchange instruments are expensive for classroom to use. At a material cost of approximately USD 62 per chamber, LeafBreath allows individual student groups to run their own light-response curves (\S~\ref{sec:app-education}) or observe a CAM leaf taking up CO$_2$ at night alongside a C3 leaf that does not (\S~\ref{sec:c3cam-validation}). Students can engage directly with net photosynthesis ($A_n$), the exact metric researchers report, rather than relying on indirect proxies. Just as Arduino \cite{arduino2026} and Micro:Bit \cite{microbit} revolutionized STEM learning for computer science and engineering, LeafBreath has the potential to transform hands-on plant biology education.

\noindent \textbf{4) Inspiring UbiComp Research on Plant Wearables and Interdisciplinary Collaboration:} While significant strides have been made in human health monitoring through mobile and wearable instruments, plant wearables have received comparatively little attention. Given the critical importance of agriculture and food security, we hope LeafBreath could provide physiological reference data for researchers developing plant monitoring and caring systems. Pairing $A_n$ records with other sensor signals could help evaluate how well those signals reflect changes in photosynthesis. These paired records could support joint studies between UbiComp researchers and plant biologists on how sensor signals relate to leaf physiology.

\section{Conclusion}

In this work, we developed and evaluated LeafBreath to make autonomous photosynthesis measurements practical on individual living leaves. Our design uses a low-cost CO$_2$ sensor and a periodically closing chamber to measure net photosynthesis without an operator at every reading. Guided by a formative study with plant researchers, we developed an adaptable chamber using 3D printing and thermoforming. We characterized the system and validated its measurements against a research-grade CIRAS-4 instrument. Our evaluations covered different leaf geometries, plant species, and experimental settings. We captured contrasting day and night CO$_2$ exchange patterns in C3 and CAM plants. In our nine-day deployment, leaves of the same plant showed different patterns of nighttime CO$_2$ uptake and recovery during drought and rewatering. These observations demonstrated the value of collecting repeated measurements from multiple leaves. We also tested the system's adaptability to small and thick leaves and demonstrated a light-response experiment suitable for a classroom. These findings show that LeafBreath can support longitudinal photosynthesis measurements on living leaves. It provides a practical basis for making leaf-level physiological data more accessible to plant researchers and educators.

\begin{acks}
We thank Rhema Amanda Phiri, Leo Zhang, Ariana Blake, Livia Ford, Sarah Jones, and Susan R. Strickler, our co-authors on the PhytoBits preprint, and the plant biology researchers who responded to our survey. We also thank Miquel Gonzalez-Meler and Ahram Cho (University of Illinois Chicago), Luke McCormack (The Morton Arboretum), and Lisa Ainsworth (University of Illinois Urbana-Champaign) for their insights on plant biology.
\end{acks}

\bibliographystyle{ACM-Reference-Format}
\bibliography{leafbreath}

\appendix
\section{Bill of Materials}
\label{app:bom}
Table~\ref{tab:bom} lists the parts of one LeafBreath chamber and what each costs.

\begin{table}[H]
  \caption{Materials cost of one LeafBreath chamber for the two controller builds (\S~\ref{sec:electronics}), at prices of 27 October 2026. Pack prices are spread over the units used, board-level parts are divided by the chambers each board runs (two for the Arduino Mega 2560 build, one for the ESP32 build). Blank: not used in that build. Tools, shipping, tax, and labor are excluded.}
  \label{tab:bom}
  \footnotesize
  \setlength{\tabcolsep}{5pt}
  \renewcommand{\arraystretch}{1.15}
  \begin{tabularx}{\linewidth}{Xrr}
    \toprule
    Component & Mega build (USD) & ESP32 build (USD) \\
    \midrule
    \multicolumn{3}{l}{\textit{Per chamber}} \\
    CO$_2$ sensor (Senseair S88 GH) & 23.44 & 23.44 \\
    Temperature and humidity sensor (SHT40 breakout) & 5.95 & 5.95 \\
    Mixing fan, 30 mm & 2.49 & 2.49 \\
    Servo (MG996R) & 4.50 & 4.50 \\
    PETG sheet, 1 mm (dome and base lid) & 1.70 & 1.70 \\
    3D-printed parts, PLA (frame, rack, pinion, clamps, molds) & 2.66 & 2.66 \\
    Silicone (Ecoflex 00-30, gaskets and lid) & 0.41 & 0.41 \\
    Hot glue & 0.14 & 0.14 \\
    Cling wrap, polyethylene (moisture barrier on the silicone lid) & $<$0.01 & $<$0.01 \\
    Copper wire, 24 AWG (ties that secure the clamp) & 0.14 & 0.14 \\
    Hook-up wire (sensor and fan wiring) & 0.70 & 0.70 \\
    \midrule
    \multicolumn{3}{l}{\textit{Per board, shared by the chambers it runs}} \\
    Fan switch MOSFET (IRLZ34N) & 0.45 & 0.90 \\
    Arduino Mega 2560 compatible board & 10.92 &  \\
    9 V, 1 A power adapter & 3.60 &  \\
    I$^2$C multiplexer (TCA9548A) & 0.58 &  \\
    Buck converter, 5 V for servos and fan (LM2596) & 0.76 &  \\
    microSD module & 0.75 & 1.50 \\
    microSD card, 8 GB & 2.59 & 5.18 \\
    ESP32-WROOM-32 development board &  & 5.32 \\
    USB cable, used as power lead &  & 3.00 \\
    USB wall charger, 5 V, 2 A &  & 3.99 \\
    \midrule
    \rowcolor{diagGreen}
    \textbf{Total per chamber} & \textbf{61.77} & \textbf{62.00} \\
    \bottomrule
  \end{tabularx}
\end{table}

\end{document}